\documentclass[manuscript]{aastex631}

\usepackage{bm}% bold math

\shorttitle{Radio wave emission by EVDF in magnetic reconnection}

\shortauthors{Yao et al., (2022)}

\usepackage{amsmath,amssymb}

\graphicspath{{./}{Figure/}}

\newcommand{\myrefeq}[1]{Eq.~(\ref{#1})}
\newcommand{\myreftab}[1]{Table~\ref{#1}}
\newcommand{\myreffig}[1]{Figure~\ref{#1}}
\newcommand{\myrefsec}[1]{Section~\ref{#1}}

\begin{document}

%%%%%%%%%%%%%%%%%%%%%%%%%%%%%%%%%%%%%%%%%%%%%%%%
% Title and authors
%%%%%%%%%%%%%%%%%%%%%%%%%%%%%%%%%%%%%%%%%%%%%%%%
\title{Wave emission of non-thermal electron beams generated by magnetic reconnection}

\correspondingauthor{Xin Yao}
\email{xin.yao@campus.tu-berlin.de}

\author[0000-0001-8988-2350]{Xin Yao}
\affiliation{Max Planck Institute for Solar System Research, 37077 {G\"{o}ttingen}, Germany}
\affiliation{Centre for Astronomy and Astrophysics, Technical University of Berlin, 10623 Berlin, Germany}

\author[0000-0002-3678-8173]{Patricio A. Mu\~{n}oz}
\affiliation{Max Planck Institute for Solar System Research, 37077 {G\"{o}ttingen}, Germany}
\affiliation{Centre for Astronomy and Astrophysics, Technical University of Berlin, 10623 Berlin, Germany}

\author[0000-0002-5700-987X]{J\"{o}rg B\"{u}chner}
\affiliation{Max Planck Institute for Solar System Research, 37077 {G\"{o}ttingen}, Germany}
\affiliation{Centre for Astronomy and Astrophysics, Technical University of Berlin, 10623 Berlin, Germany}

\author[0000-0002-4319-8083]{Jan Ben\'a\v cek}
\affiliation{Centre for Astronomy and Astrophysics, Technical University of Berlin, 10623 Berlin, Germany}

\author[0000-0001-7855-2479]{Siming Liu}
\affiliation{Purple Mountain Observatory, Chinese Academy of Sciences, 210034 Nanjing, PR China}
\affiliation{School of Physical Science and Technology, Southwest Jiaotong University, 610031 Chengdu, PR China}

\author[0000-0001-7855-2479]{Xiaowei Zhou}
\affiliation{Purple Mountain Observatory, Chinese Academy of Sciences, 210034 Nanjing, PR China}
\affiliation{Max Planck Institute for Solar System Research, 37077 {G\"{o}ttingen}, Germany}

%%%%%%%%%%%%%%%%%%%%%%%%%%%%%%%%%%%%%%%%%%%%%%%%
% Abstract
%%%%%%%%%%%%%%%%%%%%%%%%%%%%%%%%%%%%%%%%%%%%%%%%
\begin{abstract}
  \normalsize
  Magnetic reconnection in solar flares can efficiently generate non-thermal electron beams. The energetic electrons can, in turn, cause radio waves through microscopic plasma instabilities as they propagate through the ambient plasma along the magnetic field lines.
	We aim at investigating the wave emission caused by fast moving electron beams (FEBs) with characteristic non-thermal electron velocity distribution functions (EVDFs) generated by kinetic magnetic reconnection: two-streaming EVDFs along the separatrices and in the diffusion region, and perpendicular crescent-shaped EVDFs closer to the diffusion region.
	For this purpose, we utilized 2.5D fully kinetic Particle-In-Cell (PIC) code simulations in this study.
    We found that:(1) the two-streaming EVDFs plus the background ions are unstable to electron/ion (streaming) instabilities which cause ion acoustic waves and Langmuir waves due to the net current.
      This can lead to multiple harmonic plasma emission in the diffusion region and the separatrices of reconnection.
  (2) The perpendicular crescent-shaped EVDFs can cause multiple harmonic electromagnetic electron cyclotron waves through the electron cyclotron maser instabilities in the diffusion region of reconnection.
      Our results are applicable to diagnose the plasma parameters which are associated to magnetic reconnection in solar flares by means of radio waves observations.
\end{abstract}

%%%%%%%%%%%%%%%%%%%%%%%%%%%%%%%%%%%%%%%%%%%%%%%%
% Introduction
%%%%%%%%%%%%%%%%%%%%%%%%%%%%%%%%%%%%%%%%%%%%%%%%
\section{Introduction}\label{sec:introduction}
%%%%% solar radio emission
Flares are the most energetic phenomena in the solar corona. During flares, magnetic reconnection releases stored magnetic energy, and the energy can be converted into plasma bulk flow kinetic energy, heat, and acceleration of particles up to relativistic energies~\citep{Aschwanden2005,Treumann2013}. In particular, charged particles can be accelerated by reconnection, e.g., in the reconnecting current sheet~\citep[][]{Somov1997}, by the formation of cascading islands~\citep{Zhou2015,Zhou2016}, near magnetic null points~\citep[][]{Guo2010,Narukage2014,Chen2018}, in the outflow regions and separatrices~\citep[][]{Chen2015a,Liu2008,Liu2013,Glesener2012,Buchner1990}, and in underlying retracted magnetic arcades~\citep[][]{Fletcher2008}. However, it is still an open question whether the electrons are mostly generated in the central or in the outflow region of reconnection, what kind of reconnection takes place in solar eruptions and what the plasma conditions in the flaring regions are \citep[][]{Chen2015a,Chen2018}.

An important consequence of the acceleration of particles in magnetic reconnection is the formation of non-thermal electron velocity distribution functions (EVDFs)~\citep[][]{Yao2022}. Non-thermal EVDFs are prone to decay unstably~\citep{Lee2007}, which might cause non-linear phenomena like double layers~\citep{Lee2008}, anomalous transport and anisotropic heating of the coronal flare-loop plasma~\citep{Lee2010, Lee2011-1,Lee2011-2}. 
The non-thermal EVDFs can cause microscopic plasma  instabilities and generate electromagnetic emission across the entire spectrum including radio waves. Previous radio and hard X-ray observations of non-thermal radiation have confirmed the existence of such non-thermal sources of wave emissions from reconnection regions~\citep[][]{Glesener2012,Narukage2014,Fletcher2008}. For example, solar radio bursts (SRBs) are proposed to be caused by electron beams with non-thermal velocity distribution functions energized by magnetic reconnection in solar flare regions~\citep{Benz2002,Mann2009}.

In general, several kinds of waves have been associated to magnetic reconnection  \citep[see reviews in][]{Fujimoto2011, Khotyaintsev2019}.
Some of those waves can be electromagnetic and in the radio-band, which are thought to be caused by electron beams. Observations of those waves from, e.g., the flaring solar corona, can serve as a remote diagnostics tool to probe the plasma conditions of astrophysical reconnection processes.
Based on observations of coronal Type III SRBs by the upgraded Karl G. Jansky Very Large Array (VLA), \cite{Chen2018} reported waves at the fundamental plasma frequency and its harmonics, emitted from two separate source regions near magnetic null points in the solar corona.

Other waves due to reconnection have been observed in situ in the Earth's magnetosphere. Not all of them are electromagnetic and therefore can escape the plasma, but they are part of certain wave--wave interactions which can finally lead to electromagnetic radiation.
Large-amplitude and high-frequency Langmuir waves were confirmed in magnetopause reconnection \citep[][]{Burch2016a,Burch2019,Graham2018}. High-frequency electrostatic upper-hybrid (UH) waves, generated due to energetic electrons of perpendicular crescent-shaped EVDFs, were observed by the high spatial and temporal resolution multi-spacecraft Magnetospheric Multiscale (MMS) mission not only in the diffusion region of asymmetric magnetopause reconnection~\citep[][]{Graham2017} but also in central reconnection regions of the Earth's magnetotail~\citep[][]{Dokgo2019}.

%%%%% electrostatic ECWs
\cite{Menietti2002} observed electrostatic electron cyclotron waves (ECWs) due to energetic electron beams in Earth's polar regions detected by the Plasma Wave Instrument on the Polar spacecraft. 
Harmonics of electrostatic ECWs have also been observed in an over-dense plasma in the diffusion region of asymmetric reconnection in the Earth's magnetopause \citep[][]{Ronnmark1981,Tang2013,Zhou2016a,Li2020}.
By using MMS data, ~\cite{Li2020} found that the generation of multiple harmonic ECWs stands in close relation to the observation of crescent-shaped EVDFs perpendicular to the local magnetic field in the Earth's magnetopause.

%Emission mechanisms
There are several possible emission mechanisms that can be responsible for the generation of radio waves by non-thermal electrons accelerated by magnetic reconnection. 
% Plasma emission
The most widely accepted mechanism is the so-called plasma emission mechanism that relies on the wave--wave interaction of Langmuir waves induced by electron beam instabilities with ion-acoustic waves~\citep[e.g., see][and references therein]{Ginzburg1958,Melrose1970,Melrose1970a,Reid2014,Melrose2017}. 
According to the plasma emission mechanism, an electron beam first generates electrostatic Langmuir waves ($L$) via streaming-like instabilities, which are due to a positive velocity gradient in the EVDF in the direction parallel to the magnetic field: $v_{\parallel}\cdot\partial f/\partial v_{\parallel}>0$. EVDFs prone to those instabilities are known to be generated by  magnetic reconnection \citep[][]{Yao2022}.
The energy density of the Langmuir waves usually exceeds the thermal fluctuation level of the ambient plasma by several orders of magnitude.
% Note the that thermal fluctuation level can be quantified by the Klimontovich's kinetic approach, which predicts a given amount of spectral power at harmonics of the electron cyclotron frequency \citep{Yoon2017a}.
The beam-generated Langmuir waves $L$ can decay into ion-acoustic waves $S$ and fundamental emission $F$(transverse electromagnetic waves at the plasma frequency), i.e., $L\to S+F$ ~\citep[][]{Melrose1987,Melrose2017}. The fundamental emission can also be generated by a coalescence process of backward-scattered Langmuir waves $L^{\prime}$ and ion-acoustic waves $S$, i.e., $L^{\prime}+S\to F$ ~\citep[][]{Melrose1987,Melrose2017}. The (second) harmonic emission $H$ ($=H_2$), i.e., electromagnetic waves at harmonic frequencies $\omega=2\omega_{pe}$, can be generated by a coalescence of Langmuir waves $L$ and back-scattered Langmuir waves $L^{\prime}$ through $L+L^{\prime}\to H$ \citep[][]{Willes1996a,Yoon2006}. These usually circularly polarized transverse waves can escape from the ambient plasma, if their frequencies exceed the local plasma frequency which depends on the plasma density in the source region ~\citep[][]{Melrose2017,Melrose1978}. The phase velocities of these waves must also approximately exceed the speed of light, namely, they have to be superluminal.
Other mechanisms have also been recently developed to solve some problems of the standard plasma emission mechanism. For example, \citet{Che2017} proposed a wave--wave interaction process where the modulation of Langmuir waves by whistler waves during the non-linear stage of an electron two-stream instability allows a longer emission period (continuous coherent emission) than in previous models.

Not only electron beams can generate Langmuir waves but also a variety of other processes, like a current density possibly driven by the relative drift between electrons and ions, as often observed in current sheets.
A relative electron/ion streaming can offer a source of free energy for a family of instabilities collectively called electron/ion (streaming) instabilities. There are at least two well-studied instabilities.
The high-frequency one occurs when the relative streaming speed between the electron and ion populations significantly exceeds the electron thermal speed, and it is called Buneman instability \citep[][]{Buneman1958,Buneman1959}, leading to the generation of Langmuir waves \citep[][]{Gary1993,Treumann2001,Jain2011}.
The low-frequency one is called ion-acoustic instability, which requires a relative drift speed much lower than the Buneman instability, but well above the ion thermal speed and mainly occurs for $T_e \gg T_i$, leading to the generation of ion acoustic waves \citep[][]{Gary1993,Treumann2001,Hellinger2004}.
This way, a plasma featuring an electron/ion streaming could also provide the ingredients for the typical wave--wave interaction of the plasma emission mechanism to generate (fundamental) electromagnetic waves at the electron plasma frequency.

The third and higher harmonic plasma emissions, at the frequency of multiples of the local plasma frequency, were also observed in SRBs \citep[][]{Takakura1974,Reiner2019a,Cairns1986}. Various schemes have been proposed to explain the formation of higher harmonic plasma emission. \cite{Zlotnik1978} put forward that the third harmonic emission $H_3$ is generated by the coalescence process of the Langmuir waves and the harmonic emission $H$, i.e., $L+H\to H_3$. \cite{Cairns1988} generalized this idea to explain higher harmonics of transverse wave emissions, claiming that nth harmonic waves $H_n$ could be generated by a coalescence of beam-generated Langmuir waves $L$ and adjacent transverse waves $H_{n-1}$, i.e., $L+H_{n-1}\to H_n$. An alternative way of generation of harmonic transverse waves would be an interaction of Langmuir waves $L$ and harmonic electrostatic waves $L_{n-1}$, i.e., $L+L_{n-1}\to H_{n}$~\citep[][]{Gaelzer2003,Yi2007}. \citet{Ziebell2015} numerically solved the weak-turbulence equations and theoretically demonstrate the process of multiple harmonic plasma emission, including the non-linear conversion from Langmuir turbulence to electromagnetic radiation. 
\cite{Rhee2009,Ziebell2015,Yao2021} investigated both schemes of multiple-harmonic plasma emissions by means of PIC code simulations. They found that \cite{Cairns1988}'s scheme can qualitatively explain third and fourth harmonic emissions while \cite{Yi2007}'s theory is appropriate to explain simultaneously generated harmonics of electrostatic waves.
PIC code simulations of electron beams demonstrated that fundamental transverse waves at the plasma frequency can be generated because of the wave--wave interaction processes involving ion-acoustic waves~\citep{Thurgood2015,Henri2019}. The PIC simulations of ring-beam EVDFs by \citet{Zhou2020} also provided evidence for the generation of second and third harmonic emission via wave--wave interactions.
\citet{Annenkov2019a} used PIC code simulations to describe localized beams and found that emissions at harmonics of the plasma frequency can be generated due to an antenna mechanism.

% ECMI
Electron cyclotron waves (ECWs) at the relativistic electron cyclotron frequency and their harmonics have been often observed in a variety of environments. They can be generated by the so-called electron cyclotron maser instability (ECMI).
The source of free energy of this instability is due to a positive velocity gradient in EVDFs in the direction perpendicular to the local magnetic field: $\partial f/\partial v_{\perp}>0$.
ECMI could be caused by EVDFs including, e.g., cup-like distribution functions \citep{Buchner1996}, horseshoe-like EVDFs~\citep{Bingham2000,Melrose2016}, ring-~\citep{Pritchett1984,Lee2011,Yao2021} and crescent-shaped EVDFs in the velocity space perpendicular to magnetic field~\citep[][]{Burch2016,Egedal2016,Chen2016c}. 
The ECMI produces electromagnetic waves at the local electron cyclotron frequency and its harmonics as well as UH waves~\citep[][]{Benacek2019,Ni2020}. 
The electron cyclotron maser emission (ECME) mechanism predicts electromagnetic emission amplified by a quasi-linear wave--particle interaction in the magnetic field \citep[][]{Twiss1958,Melrose1984,Melrose2017}. 
Observations show that ECWs are usually X-mode polarized~\citep[][]{Ellis1962,Ellis1963,Melrose2017}. The ECME theory of X polarized ECWs at fundamental electron cyclotron frequency leads to the frequency condition $\Omega_{ce}>\omega_{pe}$ \citep[][]{Ellis1962,Ellis1963,Hewitt1985a,Melrose1984,Melrose1986,Melrose2017}, which is also necessary for electromagnetic ECWs to escape from source region. 
The above frequency condition implies a sufficiently low plasma density ($\omega_{pe}\propto \sqrt{n_e}$, with $n_e$ the electron plasma density) in regions of strong magnetic fields ($\Omega_{ce}\propto B$ with $B$ the background magnetic field strength).
In the solar corona this condition can be fulfilled in some localized regions near active regions and solar flares~\citep{Regnier2015,Morosan2016a}.

However, the plasma-to-cyclotron frequency ratio is usually the opposite one in typical space and astrophysical plasmas, in particular, in the solar corona, where the electron cyclotron frequency is often smaller than the local plasma frequency $\Omega_{ce}<\omega_{pe}$. 
On the other hand, observations of X polarized ECWs in dynamic spectra have shown that ECWs can possibly not start with the fundamental but with a higher harmonic of the electron cyclotron frequency. This implies its source region may possibly be located in a plasma where $\Omega_{ce}/\omega_{pe}<1$ \citep[][]{Treumann2011,Treumann2017}.
\citet{Treumann2017} proposed that higher harmonic of ECWs can be generated in a source region with frequency condition $\Omega_{ce}/\omega_{pe}<1$, they may escape out and be remotely observed if they are excited with high intensities. \citet{Treumann2017} pointed out that the $n=5$ harmonic ECWs can serve as cyclotron harmonic seed of the X mode polarized electromagnetic fluctuations after being amplified.
By means of PIC code simulations for the frequency condition $\Omega_{ce}/\omega_{pe}=0.1$, \citet{Ni2020} found X-O polarized fundamental and harmonic plasma emissions induced by electrostatic UH waves and other electromagnetic modes such as Z and whistler modes.

%%%%%%% EVDF simulations
A large number of numerical studies have been carried out to study the EVDFs formed by kinetic magnetic reconnection utilizing PIC code simulations \citep[see, e.g., ][and references therein]{Hoshino2001,Pritchett2004,Ng2012,Shuster2014,Munoz2016,Yao2022}. The most typical found distributions were field-aligned electron beams ~\citep{Drake2003,Pritchett2004,Che2010,Che2011}, which can cause electromagnetic wave emission.
The generation mechanisms of those (parallel) beam distributions are well understood. They are generated due to the reconnection electric field near the X-point and eventually parallel electric fields near the separatrices \citep{Treumann2013}.

Perpendicular ring- and crescent-shaped EVDFs, which can trigger the ECMI mechanism, can also be generated by reconnection. PIC code simulations revealed that crescent-shaped EVDFs can be formed by both symmetric~\citep{Shuster2014,Bessho2014} and asymmetric magnetic reconnection~\citep[][]{Hesse2014,Bessho2016,Bessho2017,Bessho2019,Shay2016,Chen2016,Price2016,Le2017}.
Perpendicular crescent-shaped EVDFs have often been detected in asymmetric magnetic reconnection through the Earth's magnetopause by the MMS mission~\citep[][]{Burch2016,Phan2016,Chen2016c,Genestreti2018,Norgren2016,Hesse2016,Egedal2016,Rager2018,Tang2019}.

Several formation mechanisms of those perpendicular crescent-shaped EVDFs have been proposed.
The redistribution of the kinetic energy of the electron motion from the direction parallel to the magnetic field into the motion in the direction perpendicular to the local magnetic field can cause the formation of the perpendicular ring- and crescent-shaped EVDFs~\citep{Voitcu2012,Voitcu2018a}. This process always takes place in kinetic magnetic reconnection since this process naturally occurs in steep magnetic gradients. \cite{Voitcu2018a} demonstrated that as electron beams propagate through a tangential magnetic discontinuity, magnetic gradient drifts can redistribute the energy of the field-aligned electron motion into that of the motion perpendicular to the local magnetic field. The formation of perpendicular crescent-shaped EVDFs in the diffusion region of reconnection can also be attributed to the meandering motion of non-gyrotropic and non-adiabatic electrons across reconnecting current sheets~\citep[][]{Hesse2014,Bessho2016,Shay2016,Lapenta2017,Bessho2019}. The exact mechanisms are still under debate, but they mostly have to do with the effects of either the $\boldsymbol{E}\times\boldsymbol{B}$ or gradient-$B$ drifts \citep[][]{Shay2016,Bessho2016,Lapenta2017}.

Many studies have been carried out to explain radio waves/emission generated by unstable EVDFs, which are not necessarily related to magnetic reconnection ~\citep{Lee2011,Ganse2012,Reid2014,Thurgood2015,Zhou2020,Yao2021}. There have also been studies analyzing radio wave emissions due to magnetic reconnection without focusing on specific features of the EVDFs caused by reconnection or their associated kinetic plasma instabilities~\citep{Sakai2005,Karlicky2007}.
A direct relation of the generation of electromagnetic waves to non-thermal EVDFs generated by kinetic magnetic reconnection is, to the best of our knowledge, still missing. In this paper, we intend to bridge this gap.

Our approach is to directly use the non-thermal EVDFs generated by numerical simulation of kinetic magnetic reconnection, which feature sources of free energy for micro-instabilities, as initial conditions of beam-plasma simulations with the objective to investigate their consequent radio wave emission. This two-step approach allows a higher frequency and wavenumber resolution than that of the immediate diagnostics of radio waves in reconnection simulations. This approach provides thus a better understanding not only of the instabilities caused by non-thermal EVDFs but also of the properties of the resulting radio waves.

For this purpose, we used the results of PIC code simulations of 3D kinetic magnetic reconnection performed in a previous study~\citep{Yao2022}. They identified non-thermal EVDFs as potential sources of free energy for the generation of radio waves in those reconnection simulations  based on an unsupervised machine learning technique \cite[][]{Dupuis2020}. According to~\cite{Yao2022}, two characteristic EVDFs that are prone to instabilities relevant to radio emission were identified: two-streaming EVDFs near the separatrices and diffusion region, while  perpendicular crescent-shaped EVDFs in the diffusion and the outflow region of reconnection.
We used those EVDFs in a parametrized way as initial conditions for separate beam-plasma simulations.

Previous works have also extracted and parametrized EVDFs from reconnection simulation to assess their stability properties via independent 2D simulations.
This is because a high frequency and wavenumber resolution are needed to resolve the unstable waves, which is usually not feasible in 3D reconnection simulations.
The usual method is to simulate that parametrized EVDF in a homogeneous background plasma in a 2D or 1D geometry. Indeed, \citet{Goldman2008} carried out 1D and 2D kinetic Vlasov simulations to understand the electron holes caused by the Buneman instability, which is triggered by a EVDF taken from 2D PIC simulations of magnetic reconnection. They used a spatially varying parallel velocity profile to mimic the current sheet width but otherwise constant density.
\citet{Divin2012} simulated separatrix instabilities driven by unstable EVDFs from 2D PIC simulations of magnetic reconnection with realistic mass ratio. A step forward of this work is the inclusion of the force balance at the separatrices into the initialization of the EVDF simulations. They neglected, however, gradients parallel to the magnetic field.  Note that none of those works aimed at the analysis of the resulting radio emission or velocity distribution functions (VDFs) prone to the electron cyclotron maser instability, so our work attempts to bridge that gap. \cite{Hesse2018} numerically solved the electrostatic dispersion relations of typical EVDFs found in the separatrices of reconnection, with the purpose to assess their stability and role in generating the resulting electrostatic soliton-like waves.

The main objective of this paper is thus to relate the non-thermal EVDFs caused by magnetic reconnection with the resulting electromagnetic radiation due to the kinetic instabilities triggered by those EVDFs and also the ion distribution functions.
This is of fundamental importance to understand, for example, the plasma properties of solar coronal reconnection based on the observed radio waves caused by the accelerated electron beams.
We address only the initial stage of this process at its  source region; a more complete model would require to take into account the transport of electrons as they propagate through the solar corona and solar wind.

The content of this paper is organized as follows: in \myrefsec{sec:model} we first describe the numerical simulations of 3D kinetic magnetic reconnection and the resulting characteristic non-thermal EVDFs. We then describe the numerical setup of simulations with a few of those non-thermal EVDFs as initial conditions. In \myrefsec{sec:results} we presented our analysis of the resulting radio waves by these unstable EVDFs. Our conclusions are summarized in \myrefsec{sec:conclusions}.

%%%%%%%%%%%%%%%%%%%%%%%%%%%%%%%%%%%%%%%%%%%%%%%%
% Numerical Model
%%%%%%%%%%%%%%%%%%%%%%%%%%%%%%%%%%%%%%%%%%%%%%%%
\section{Numerical Model}\label{sec:model}
%%%%%%%%%%Reconnection simulations and EVDFs

\subsection{3D magnetic reconnection simulation}

We first briefly summarized the 3D magnetic reconnection simulation \citep[][]{Yao2022}, from which we obtained the EVDFs to be analyzed in this study. Those simulations were performed with the fully-kinetic 3D Particle-in-Cell (PIC) code ACRONYM~\citep[][]{Kilian2012}, which is appropriate to model the collisionless plasmas of the solar corona as it solves the fully-kinetic Vlasov-Maxwell system of equations.

The initial equilibrium was a double Harris current sheet equilibrium without an external guide field (also known as antiparallel reconnection).
Both electrons and ions are initialized following this equilibrium. From now on we will denote those species as Harris population. In addition, we also impose a homogeneous background consisting also of ions and electrons with the same temperature as the Harris population and with a density 0.1625 times that of the peak Harris current sheet density.

A small perturbation was applied to the equilibrium magnetic field in order to accelerate the reconnection onset. The number of grid points of the simulation box was $256\times 512\times 1024$, and the physical simulation domain was $4 d_i\times8d_i\times 16d_i$, where $d_i$ is the ion skin depth.
We apply periodic boundary conditions in all directions. The ion-to-electron mass ratio was $m_i/m_e=100$. Other parameters of this simulation can be seen in the central column of Table~\ref{tab:PIC_Simulations_Parameters_3D_2D}.

As described in \citet{Yao2022}, the evolution of this magnetic reconnection simulation is characterized by reconnection rates that keep increasing almost until the end of the simulated time. 
\textbf{The same can be observed by means of other diagnostics, like the root mean square (RMS) values of the magnetic field, magnetic energy, etc.}
Our system does not reach a steady state.
The reconnection rates reach a value of 0.1 in normalized units (i.e., $v_AB_{\infty}/c$ in CGS units) at $t=5.25\ \Omega_{ci}^{-1}$.
This is the standard value for fast reconnection \citep[][]{Cassak2017} \textbf{as well as the value reported by some turbulent reconnection simulations  \citep[][]{Wendel2013,Liu2018}}, although it can be higher in some cases, specially when Buneman-like turbulence is triggered in magnetic reconnection \citep{Che2011,Che2017a, Munoz2018a}.
\textbf{Simulations of kinetic turbulence where many sites of magnetic reconnection occur show a distribution of reconnection rates below and above 0.1 \citep{Haggerty2017}}.
In our case, the time after the reconnection rates reach values higher than 0.1 is dominated by unphysical effects, which are the interaction with the second current sheet in magnetic reconnection (because of periodic boundary conditions) and the depletion of the available magnetic flux (which is due to the reduced simulation box size).
Previous 3D magnetic reconnection simulations also usually focused on diagnostics at times when the reconnection rates are near 0.1 \citep{Che2011,Munoz2018a}. In this study, we decided to analyzed EVDFs at $t=5.25\ \Omega_{ci}^{-1}$ when the reconnection rates reach 0.1, same as the analysis of \citet[][]{Yao2022}.

\begin{deluxetable*}{ccc}
  \tablewidth{0pt}
  \tablenum{1}
  \tablecaption{A comparison of parameters of the 3D simulations of magnetic reconnection and 2.5D simulations of a beam-plasma system. For reconnection simulations, some parameters have different values depending on their distance from the current sheet center (as per the Harris current sheet equilibrium). In this case, the first indicated value is the asymptotic value, while the values to the right of the arrows are calculated at the current sheet center. Here $\lambda_D$ is the Debye length,  $d_i$ the ion skin depth, $\omega_{pe}$ the electron plasma frequency calculated at the current sheet center.  $\Omega_{ce}$ and $\Omega_{ci}$ are the electron and ion cyclotron frequencies calculated asymptotically away from the current sheet center, respectively. $v_A$ is the Alfv\'en speed, $\beta_e$ the electron plasma beta,  $n_e$ the electron number density, $\Delta x$ the grid cell size, $\Delta t$ the timestep. \label{tab:PIC_Simulations_Parameters_3D_2D}}
  \tablehead{
  \colhead{Parameter} & \colhead{3D magnetic reconnection}   & \colhead{2.5D beam-plasma system}
  }
  \startdata
  grid points                        & $256\times 512\times 1024$              & $2048\times 2048$ \\
  physical size                      & $4\times 8\times 16\;d_i^3$             & $14.3 \times 14.3\;d_i^2$ \\
  $\Delta x\ [\lambda_{De}]$         & 1.56                                    & 1\\
  $m_i/m_e$                          & 100                                     & 100 \\
  $v_{th,e}/c$                       & 0.1                                     & 0.07 \\
  $T_e\, [eV]$                       & 5110                                    & 2503  \\
  $T_i/T_e$                          & 1                                       & 1\\
  $v_A/c$                            & 0.05 $\to$ 0                            & 0.045 \\
  $\beta_e$                          & 0.08 $\to$ $\infty$                     & 0.047 \\
  $\omega_{pe}/\Omega_{ce}$          & 2$\to\infty$                            & 2.2\\
  $n_e\,[{\rm cm}^{-3}]$             & $1.3\times10^9 \to (7.9+1.3)\times10^9$ & $(7.9+3.95)\times10^9$  \\
  Particles per cell                 & 13 $\to$ (80+13)                        & 100+50  \\
  $\Delta t\ [\omega_{pe}^{-1}]$     & 0.087                                   & 0.035\\
  Total time $[\Omega_{ci}^{-1}]$    & 9                                       & 3.6 \\
  Output period $[\omega_{pe}^{-1}]$ & 55.5                                    & 0.35 \\
 \enddata
\end{deluxetable*}

\begin{figure}[ht!]
  \plotone{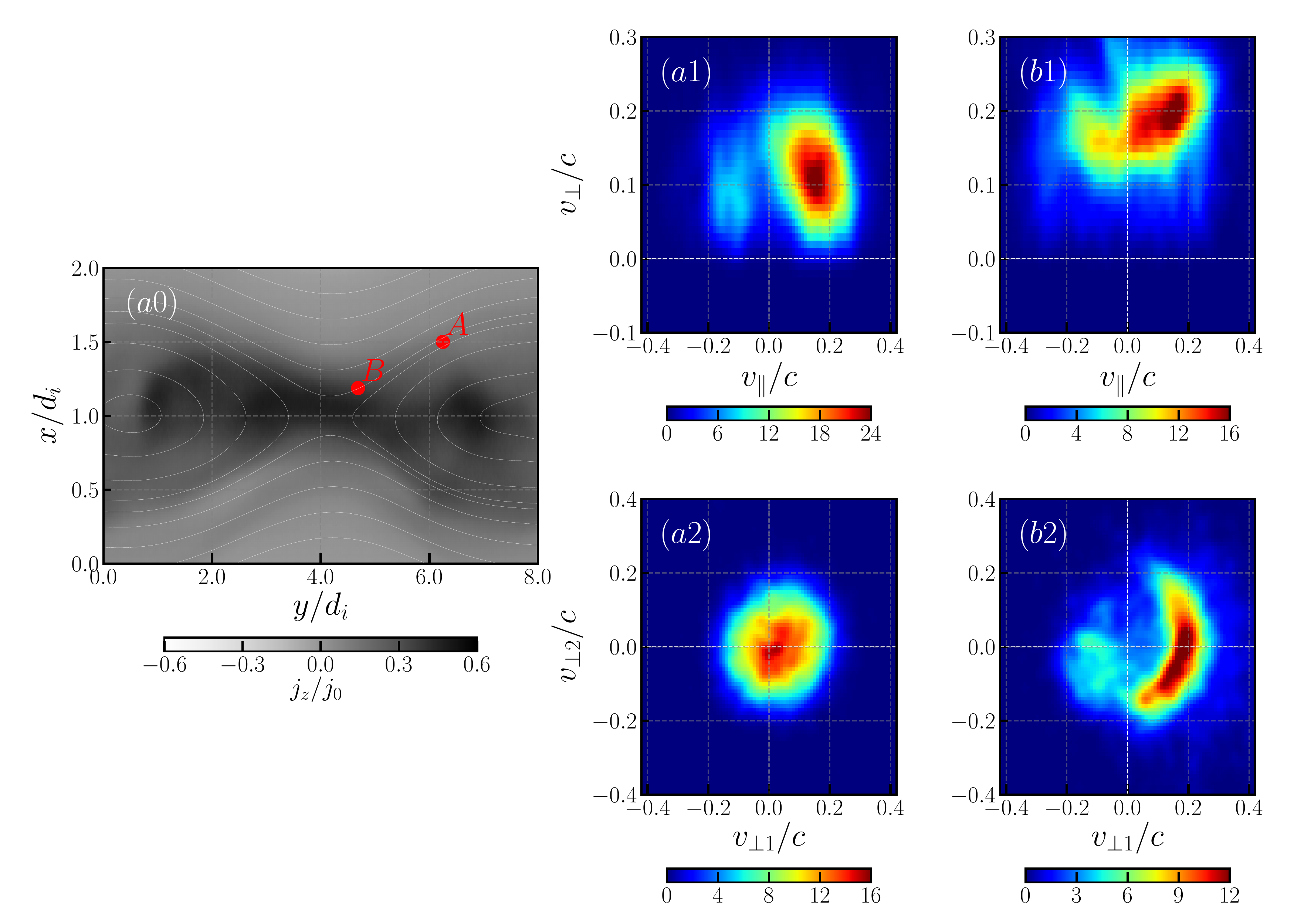}
  \caption{Non-thermal EVDFs generated by 3D kinetic simulation of antiparallel magnetic reconnection on the plane $z=8.0d_i$ during time $t=5.25\ \Omega_{ci}^{-1}$ (see details in \citep{Yao2022}). (a0) Normalized current density $j_z$ on the $x-y$ reconnection plane $z=8\ d_i$ (with $z$ along the out-of-plane direction) and here $j_0=en_0v_{the}$. Magnetic field lines (white curves) on the reconnection plane are overlaid. (a1--a2): The two-streaming EVDF generated in the separatrix region (point A) and (b1--b2): the perpendicular crescent-shaped EVDF generated near diffusion region (point B) in the $v_{\parallel}-v_{\perp}$ and $v_{\perp1}-v_{\perp2}$ planes separately. The parallel-$\parallel$ direction is along the local magnetic field. The EVDFs are estimated as the normalized probability density function of electrons of the Harris population in the velocity space.
  \label{fig:EVDF_MR3D}}
\end{figure}

In this study, we choose two characteristic EVDFs formed at two different locations on a given reconnection plane. The locations are denoted by the points A and B (see red dots) in \myreffig{fig:EVDF_MR3D}(a0).
The characteristic non-thermal EVDFs for FEBs at these locations are: (A) two-streaming EVDF generated in the separatrix region (see \myreffig{fig:EVDF_MR3D}(a1--a2)) and (B) perpendicular crescent-shaped EVDF formed near the diffusion region (see \myreffig{fig:EVDF_MR3D}(b1--b2)).
Those EVDFs were calculated with the electrons that belong to the so-called
Harris population (which initially establishes the initial equilibrium) which, for the purposes of the present work, can be considered as the electron beams.
They can provide the source of free energy for electron/electron streaming instabilities depending on their positive velocity gradients, and also for electron/ion streaming instabilities which depend on the relative drift speed between the bulk flow speed of the entire EVDF and the ion distribution.
A positive  velocity gradient in the parallel EVDFs is a necessary condition for the streaming instabilities,  eventually causing Langmuir waves due to inverse Landau damping, as per the so-called Penrose criterion \citep{Penrose1960}. The rigorous necessary and sufficient conditions requires to not only determine the regions with positive velocity gradients, but also to take into account any negative contribution to the instability from the rest of the distribution function.
%A positive  velocity gradient in the parallel ion distribution function with respect to the EVDF is also a necessary condition for electron/ion instabilities which can cause both high frequency waves such as a Langmuir waves (due to the Buneman instability), but also low frequency waves, such as ion acoustic waves (due to the ion-acoustic instability).

There is also a background population not shown in \myreffig{fig:EVDF_MR3D}, which tends to reduce positive velocity gradient(s) due to the beam population and thus suppress streaming-like instabilities. However, the electron beam can propagate through a background with different densities and so this effect will vary point to point within the reconnection region, so we focus mainly on the beam population.

In this 3D antiparallel magnetic reconnection simulation, the magnetic field lines roughly are onto the reconnection plane. Electron beams then move along the magnetic field lines (see white curves in \myreffig{fig:EVDF_MR3D}(a0)) from the X-point toward the separatrices.

Note that the third dimension of this magnetic reconnection study (in contrast to other 2D reconnection simulations) allows the release of energy of unstable distributions in this direction.
More precisely, field-aligned beams along the third direction will generate unstable waves with a wavenumber in this direction. In a pure 2D configuration that process is prohibited, and therefore, a distribution that could be unstable in 3D will be stable in a 2D geometry. As a result, there should be very different waves due to those unstable EVDFs in a 3D simulation in comparison to its 2D counterpart. The unstable EVDFs in 3D magnetic reconnection are observed for relatively long time-scales (i.e. ion time-scales), and thus they are always continuously generated even though their free energy should be depleted due to streaming-like instabilities.

For further details about this simulation and a more comprehensive analysis of EVDFs, see \citep[][]{Yao2022}.

\subsection{2.5D beam-plasma simulations}
% Our simulation
We used two characteristic EVDFs described above as initial conditions for independent simulations performed with the 2.5D version of ACRONYM, i.e., a 2D mesh grid in space and the full 3D velocity coordinates.
We considered an electron-ion plasma with an ion-to-electron mass ratio $m_p/m_e=100$, same as the reconnection simulations.
The plasma consists of three species of particles: background electrons and ions, and beam electrons streaming at a given drift speed.
The initial electron plasma frequency is set to be $\omega_{pe}=5.0\times 10^9\ rad/s$, which corresponds to an electron number density of $n_0=7.9\times 10^9\ cm^{-3}$, typical for the solar corona~\citep[][]{Aschwanden2005}, and also the same values as the reconnection simulations.
The ratio of the electron cyclotron frequency to the electron plasma frequency is $\Omega_{ce}/\omega_{pe}=0.45$.
Such frequency ratio is generally found in the separatrices in our 3D kinetic collisionless magnetic reconnection simulation. It can also be found at some locations in the solar corona ~\citep[][]{Morosan2016a}.
This implies an electron plasma beta of $\beta_e=0.047$.
The size of the simulation box is $(L_x,L_y)=(N_x,N_y)\times \Delta x= (2048,2048)\times\Delta x$ along the $x$ and $y$ directions, respectively. We set the grid cell size to be $\Delta x=\lambda_D$ (with the Debye length of $\lambda_D=0.42$~cm).
In order to satisfy the Courant-Friedrichs-Lewy (CFL) condition, we imposed the condition $c\Delta t/\Delta x=1/2<1/\sqrt{2}$.
Periodic boundary conditions are applied in both directions of the simulation box.
The time step is $\Delta t=0.035\ \omega_{pe}^{-1}$.
The background magnetic field is assumed to be constant throughout the box, and the direction of the magnetic field defines the $x$ direction of the simulation box, namely, $\vec{B}_0=B_0\vec{e}_x$.
From here on, we refer to the $x$ direction as the parallel direction.
We set the thermal speed of the background electron plasma as $v_{the}=0.07c$ and of beam electrons $v_{th,bm}=0.08c$.
This beam thermal speed was calculated by fitting the parallel electron distribution function shown in \myreffig{fig:EVDF_MR3D} (a1) and determining its standard deviation.
The perpendicular distribution exhibits similar values.
Because of the strong non-Maxwellian features of the crescent-shaped EVDF shown in Figure (b1-b2), a Maxwellian fitting is not meaningful, so we choose to use the same temperature values as those determined for the point A.
As for the background population, not shown in \myreffig{fig:EVDF_MR3D}, we used the thermal speed values resulting from the Maxwellian fitting carried out in Figure 8 (b2) of \citet[][]{Yao2022}.
Nevertheless, all the results obtained in this work are relatively insensitive to small variations in the thermal speed (or temperature) within the range $v_{the}/c=0.05\sim 0.1$.

\myreftab{tab:PIC_Simulations_Parameters_3D_2D} shows a comparison between the parameters of the 3D simulation of magnetic reconnection and those of the 2.5D simulation of beam-plasma system.
In contrast to magnetic reconnection simulations, these beam-plasma simulations allow to reach a higher frequency and wavenumber resolution and provide a better understanding of the stability properties and the resulting radio waves. For 2.5D simulation of beam-plasma systems, the size of the simulation box allows a wavenumber resolution of $\Delta k_x=\Delta k_y=0.044\ \omega_{pe}/c$. Based on a sampling period of $10\Delta t$, the simulation allows to obtain a frequency up to $\omega_{max}\approx 3\pi\omega_{pe}$ in the frequency domain.
The presented dispersion relation analysis is based on a time window which allows a frequency resolution of $\Delta \omega=0.035\ \omega_{pe}$ (by 512 samplings) or even higher frequency resolution such as $\Delta \omega=0.0175\ \omega_{pe}$ (by 1024 samplings) or $\Delta \omega=0.0044\ \omega_{pe}$ (by 4096 samplings).  For the 3D simulation of magnetic reconnection, the size of the simulation box allows a wavenumber resolution of $\Delta k_x=0.08\omega_{pe}/c,\Delta k_y=0.04\omega_{pe}/c,\Delta k_z=0.02\omega_{pe}/c$.
However, the 3D simulation allows maximum frequency only up to $\omega_{max}=0.025\omega_{pe}$, which is not enough to analyze the high frequency waves such as Langmuir waves and electron cyclotron waves caused by micro-instabilities. Due to the large output period comparable to the reconnection time scale at ion cyclotron time (i.e., $1429\Delta t=0.025\Omega_{ci}^{-1}$), only a few outputs are available.
This is because of computational limitations: the parallel output (in HDF5 format) of those simulations that utilize thousands of cores tends to slow down the entire simulation in most supercomputers file systems, so that a too often output significantly hinders the speed of the code.
Therefore, a 2.5D simulation with high frequency and wavenumber resolutions is necessary to investigate the waves caused by the relevant micro-instabilities that are studied here.

%%%%%%%%%% macro-particles
For the simulations in this study, we choose $N_{bg}=100$ macro-particles per cell for the background plasma and  $N_{bm}=50$ macro-particles per cell for the electron beam. This represents a beam-to-background density ratio of $n_{bm}/n_{bg}=1/2$ since the number of macro-particles per cell is proportional to the physical number density, provided a constant ratio of macro-particles to physical particles.
Such a relatively high beam-to-background density ratio can be generated in a local region in the above described 3D magnetic reconnection simulation (see locations denoted by solid red circles in \myreffig{fig:EVDF_MR3D}(a0)). Several examples of distributions with this beam-to-background density ratio can be found in \citet[][]{Yao2022}.
% In fact, the beam-to-background density ratio as high as $N_{bm}/N_{bg}\ge1/2$ can be locally produced elsewhere i
% As discussed in the \myrefsec{sec:results}, we found that the excitation of transverse emission by two-streaming instabilities is sensitive to the beam-to-background density ratio and the parallel drift speed (i.e., averaged kinetic bulk flow speed) of the electron beams.
We noted that our convergence tests of similar beam-plasma simulations verified that simulations with a larger number of macro-particles per cell for both background and beam plasma basically produce similar results \citep[][]{Yao2021}. In order to reduce the level of numerical noise, a second order shape function was used.

%%%%%%%%%% velocity
From here on we used the term ``momentum'' for the momentum per unit mass, which is equivalent to the relativistic velocity in its four-vector form, namely, $p_{\parallel}=v_{\parallel}$, $p_{\perp}= v_{\perp}$. 
Note that the four-velocities $v_{\parallel}$ and $v_{\perp}$ differ from the ordinary three-velocity by a factor of $\gamma$, where the Lorentz factor is $\gamma=\sqrt{1+(v_{\parallel}^2+v_{\perp}^2)/c^2}$. The particle kinetic energy is $E_k=(\gamma-1)m_ec^2$.

For the non-thermal EVDFs generated in our kinetic magnetic reconnection simulation, the averaged kinetic bulk flow speed (or drift speed) of beam electrons along the direction of the local magnetic field (i.e., the parallel drift speed) is $|u_{d\parallel}|<0.2c$. The maximum value $0.2c$ corresponds to the typical drifts speeds obtained by fitting Maxwellian distributions to our two selected EVDFs found in the separatrices and the diffusion region (see \myreffig{fig:EVDF_MR3D}(a1,b1)). 
We also found the excitation of harmonic(s) of plasma emission is sensitively dependent on the parallel drift speed of electron beam and the beam-to-background number density ratio (this will be discussed below).
Therefore our beam-plasma simulations are initialized with a beam drift speed of $u_{d\parallel}=0.2c$.
In observations, the drift speed of electron beams is found to be either non-relativistic, $0.2-0.5c$ \citep[][]{Wild1959,Alvarez1973} or mildly relativistic with $>0.6c$ \citep[][]{Poquerusse1994a,Klassen2003}. The drift speed of the electron beams can be deduced from observations of Type III radio bursts as summarized in \citet[][]{Reid2014,Reid2018}.
% We found the excitation of harmonic(s) of plasma emission is sensitively dependent on the parallel drift speed of elelctron beam and the beam-to-beanckground number density.
% In this study, we chose $u_{d\parallel}=0.2c$ mainly considering the drift speed of elelctron beams generated by magnetic reconnection as mentioned above and the aim to generate multiple harmonic plasma emission.
% In fact, for an electron beam with drift speed $u_{d\parallel}=0.2c$, if the beam-to-background number density is high enough, e.g., $n_{bm}/n_{bg}\ge1/2$, multiple harmonic plasma emission can be generated by the two-streaming instabilities (see a discussion in \myrefsec{sec:Langmuir_waves}).
Meanwhile, the kinetic bulk flow speed perpendicular to the direction of the local magnetic field (i.e., the perpendicular drift speed) of the electron beam with perpendicular crescent-shaped EVDF is set to the same value obtained by the fitting of the EVDFs from \myreffig{fig:EVDF_MR3D}(b1,b2), i.e., $u_{d\perp}=0.2c$.
Both parallel and perpendicular drift speeds $u_{d\parallel}=u_{d\perp}=0.2\ c$ correspond to a Lorentz factor $\gamma=1.02$ and kinetic energy $E_k=10.13\ keV$.

We conducted three simulation runs that are summarized in the \myreftab{tab:PIC_Simulations_Parameters_3runs}.
Among them, Run1 represents a control case with only background plasma.
This thermal and homogeneous plasma can excite most of the normal plasma modes (surfaces or curves in the frequency--wavenumber space) due to the fluctuations caused by the thermal noise. In a PIC simulation the physical thermal noise is provided by the macro-particles, where each of them represents a bulk of physical particles. Therefore the thermal noise level in a PIC simulation is higher than in a real plasma (see, e.g., section 5.3 of \citet{Melzani2013}). In either case the normal plasma modes are excited above the thermal level, but only in the Fourier space  (frequency-wavenumber) region where they are either undamped or weakly damped.
The excitation of all plasma normal modes is routinely used as a test problem to verify the correctness of PIC-code algorithms. For example, \citet{Kilian2017} showed the presence of most plasma modes  in simulations of a homogeneous thermal plasma (i.e., without sources of free energy) with a variety of kinetic codes and plasma models, including a fully-kinetic PIC code.\\

This way, waves are excited due to the electron beams in Run2 and Run3, which mostly (but not always) occur onto the surfaces or curves (in the frequency-wavenumber space) of the normal plasma modes,
can be compared to the waves excited by thermal fluctuations onto those same normal modes in Run1.
Note that any wave excited by a plasma instability, like in our Runs2 and Run3, features a spectral power that is orders of magnitude larger than the normal plasma modes excited by the thermal noise in Run1. So the waves by those different processes can be easily distinguished.

\begin{deluxetable*}{cccccc}
  \tablewidth{0pt}
  \tablenum{2}
  \tablecaption{Parameters for the electron velocity distribution function (EVDF) of the electron beams for the simulation runs. \label{tab:PIC_Simulations_Parameters_3runs}}
  \tablehead{
    \colhead{Run} &\colhead{Background} &\multicolumn{4}{c}{Beam}\\
    \colhead{}    &$v_{the}$            &\colhead{$v_{th,bm}$}  &\colhead{$u_{d\parallel}$} &\colhead{$u_{d\perp}$} & \colhead{EVDF}
  }
  \startdata
  1 &  $0.07c$ & -       & -       & -       & no beam\\
  2 &  $0.07c$ & $0.08c$ & $0.2c$ & $0$     & two-streaming\\
  3 &  $0.07c$ & $0.08c$ & $0.2c$ & $0.2c$  & drifting crescent-shaped\\
  \enddata
\end{deluxetable*}

\begin{figure}[ht!]
  \plotone{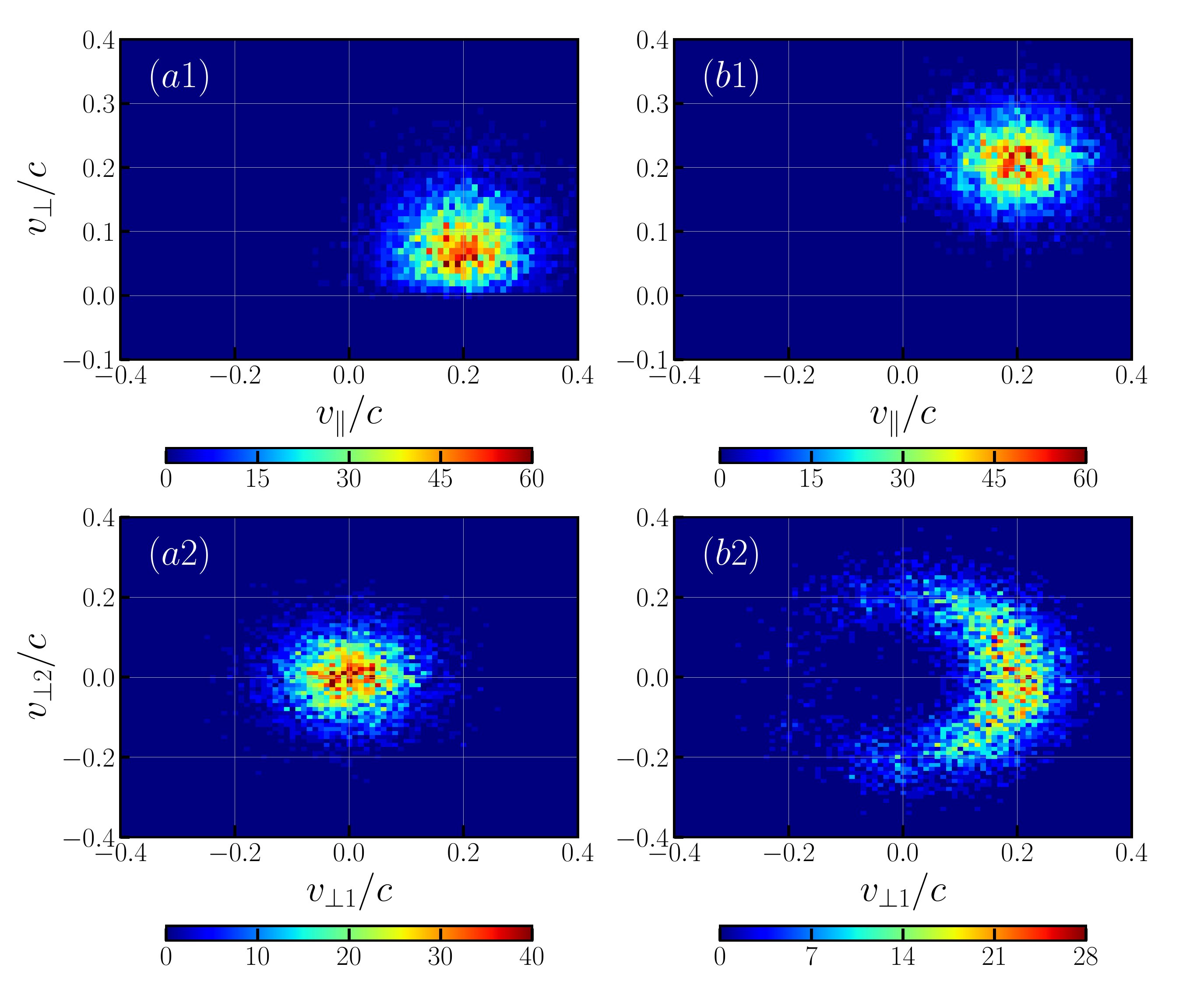}
  \caption{Initial EVDFs of our beam-plasma simulations parametrized from the reconnection EVDFs in \myreffig{fig:EVDF_MR3D}. Two-streaming (a1--a2) and perpendicular crescent-shaped (b1--b2) EVDFs in the $v_{\parallel}-v_{\perp}$ and $v_{\perp1}-v_{\perp2}$ planes, respectively. (a1--a2) is used to initialize Run2, while (b1--b2) is used to initialize Run3. The two-streaming EVDF is generated by the product of \myrefeq{eq:PDFvparaBeam_Maxw} and \myrefeq{eq:PDFvperpBeam_Maxw} and the perpendicular crescent-shaped EVDF by the product of \myrefeq{eq:PDFvparaBeam_crec} and \myrefeq{eq:PDFvperpBeam_crec}. See details about the specific parameters and the fitting of the reconnection simulations in the text.
  \label{fig:EVDF_Yao}}
\end{figure}

We initialized the beam-plasma system by prescribing a particle distribution function $f(\vec{x},\vec{v})$ in the phase space $(x,y)\times(v_{\parallel},v_{\perp})$.
The distribution is spatially homogeneously distributed and thus independent on the $(x,y)$ coordinates.
The background particle distribution function is expressed as follows,
\begin{equation}
    f_{bg}\left(x,y,v_{\parallel},v_{\perp}\right)=n_{0,bg}f_{\parallel}\left(v_{\parallel};v_{the}\right)f_{\perp}\left(v_{\perp};v_{the}\right)\label{eq:PDFxvparavperpBackground}
\end{equation}
where
\begin{equation}
    f_{\parallel}\left(v_{\parallel};v_{the}\right)=\frac{1}{\sqrt{2\pi v_{the}^2}}\exp\left(-\frac{v_{\parallel}^2}{2v_{the}^2}\right)\label{eq:PDFvparaBackground}
\end{equation}
\begin{equation}
  f_{\perp}\left(v_{\perp};v_{the}\right)=\frac{1}{v_{the}^2} v_{\perp}\cdot\exp\left(-\frac{v_{\perp}^2}{2v_{the}^2}\right)\label{eq:PDFvperpBackground},
\end{equation}
here $n_{0,bg}=N_{bg}(M/\Delta V)=n_0$ is the background number density, $N_{bg}$ is number of macro-particles per cell of the background plasma, $M$ is the ratio of physical to numerical particles and $\Delta V$ is the cell volume, $v_{the}$ is the thermal speed of the background electrons.

The distribution function of the beam electrons is expressed in the following form:
\begin{equation}
    f_{bm}\left(x,y,v_{\parallel},v_{\perp}\right)=n_{0,bm}\cdot f_{\parallel}(v_{\parallel};u_{d\parallel},v_{th\parallel})f_{\perp}(v_{\perp};u_{d\perp},v_{th\perp})\label{eq:PDFxvparavperpBeam},
\end{equation}
here $n_{0,bm}=N_{bm}(M/\Delta V)$ is the beam number density, $N_{bm}$ is number of beam macro-electrons per cell.

The two kinds of EVDFs of FEB for Run2 and Run3 are:
\begin{enumerate}
  \item  two-streaming EVDF
  \begin{equation}
    f_{\parallel}(v_{\parallel};u_{d\parallel},v_{th\parallel})=\frac{1}{\sqrt{2\pi v_{th\parallel}^2}}\exp\left[-\frac{\left(v_{\parallel}-u_{d\parallel}\right)^2}{2v_{th\parallel}^2}\right]\label{eq:PDFvparaBeam_Maxw}
  \end{equation}
  \begin{equation}
    f_{\perp}(v_{\perp};u_{d\perp},v_{th\perp})=\frac{1}{v_{th\perp}^2} v_{\perp}\cdot\exp\left[-\frac{v_{\perp}^2}{2v_{th\perp}^2}\right]\label{eq:PDFvperpBeam_Maxw}
  \end{equation}
  where $v_{th\parallel}$ and $v_{th\perp}$ are the thermal speeds along and perpendicular to the direction of the local magnetic field. We consider here only the case of an initially isotropic beam plasma, i.e, $v_{th\parallel}=v_{th\perp}=v_{th,bm}$.
  \item perpendicular crescent-shaped EVDF
  \begin{equation}
    f_{\parallel}(v_{\parallel};u_{d\parallel},v_{th\parallel})=\frac{1}{\sqrt{2\pi v_{th\parallel}^2}}\exp\left[-\frac{\left(v_{\parallel}-u_{d\parallel}\right)^2}{2v_{th\parallel}^2}\right]\label{eq:PDFvparaBeam_crec}
  \end{equation}
  \begin{equation}
    f_{\perp}(v_{\perp};u_{d\perp},v_{th\perp})=\frac{1}{N}v_{\perp} \cdot \exp\left[-\frac{(v_{\perp}-u_{\perp0})^2}{2v_{th\perp}^2}-\frac{(\phi-\phi_0)^2}{2\phi_{th}^2}\right]\label{eq:PDFvperpBeam_crec}
  \end{equation}
  \begin{equation}
    N=\sqrt{\pi}\beta\cdot \mathrm{erf}(\pi/\beta)\cdot\left[\frac{\sqrt{\pi}}{2}\alpha v_0\left[\mathrm{erf}(v_0/\alpha)+1\right]+\frac{\alpha^2}{2}e^{-v_0^2/\alpha^2}\right]
  \end{equation}
  here $\phi=\arctan (v_{\perp2}/v_{\perp1})$ is the polar angle with the two orthogonal components of the perpendicular speed $v_{\perp1}$ and $v_{\perp2}$, and $v_{\perp}=\sqrt{v_{\perp1}^2+v_{\perp2}^2}$. Here $\phi_{th}$ determines the angular width spread of velocity distribution function about the angle $\phi_0$ in the perpendicular velocity space $v_{\perp1}-v_{\perp2}$.
  The quantity $N$ is the normalization factor and $\alpha=\sqrt{2}v_{th\perp},\ \beta=\sqrt{2}\phi_{th}$. In our simulations, $\phi_0=0$ and $\phi_{th}=0.6\pi$, which empirically correspond to the typical perpendicular crescent-shaped EVDF found in diffusion region (see \myreffig{fig:EVDF_MR3D}(b2)). Note that our simulations are implemented in Cartesian coordinates: the parallel direction is field-aligned as mentioned above, thus $v_{\parallel}=v_x$. For the components of perpendicular velocity, we simply assign $v_{\perp1}=v_y$ and $v_{\perp2}=v_z$.
\end{enumerate}

Both beam electrons, with the two-streaming and perpendicular crescent-shaped EVDFs, are initialized by using \myrefeq{eq:PDFxvparavperpBeam} (\myreffig{fig:EVDF_Yao}).
Note that both of these non-thermal EVDFs for FEBs have the two-streaming EVDF along the parallel direction in velocity space (see \myreffig{fig:EVDF_Yao}(a1,b1) in $v_{\parallel}-v_{\perp}$ plane) because fast moving electron beams inherently have a positive parallel drift speed (bulk flow kinetic speed), namely, $u_{d\parallel}>0$.

The main three beam-plasma simulations to be investigated in this paper and the associated kinetic processes that are expected in them can be summarized as follows:
\begin{itemize}
    \item Run1: There is no beam at all. This is used as a control case. Normal plasma wave modes are generated as a result of the background thermal plasma noise (See response to the Remark 3 above). Those visible waves include: Langmuir waves, ordinary (O mode), extraordinary (X mode) waves, whistler waves, and electrostatic Bernstein waves. Note that they have small amplitudes in comparison with the waves generated by instabilities. Waves subject to kinetic damping, such as Langmuir waves, are only visible for large wavelengths, where their damping is weak.
    \item Run2: We use an electron beam featuring a streaming-like EVDF. The kinetic processes are dominated by instabilities due to the field-aligned Maxwellian electron beam, which in general leads to a relative electron/ion streaming.
    This relative streaming causes a net current and thus oscillations at the electron plasma frequency. Because of thermal effects, waves at those frequencies experience dispersion and so they are commonly called Langmuir waves.
    Later during the evolution of the system the same electron/ion streaming leads to an electron/ion streaming instability, causing ion-acoustic waves at much lower frequencies.
    The existing Langmuir waves can then interact with the generated ion-acoustic waves through a wave--wave interaction process known as the plasma emission mechanism. The resulting electromagnetic waves at the electron plasma frequency (fundamental) and their harmonics can escape the plasma and be remotely observed.

    \item Run3: We use an electron beam featuring a crescent-shaped EVDF. This beam is also characterized by the same drift speed in the parallel direction to the local magnetic field as the EVDF for Run2. Therefore all kinetic processes taking place in Run2 also occur in this simulation. In addition, this crescent-shaped EVDF features a positive velocity gradient in the direction perpendicular to the local magnetic field.
    The crescent-shaped EVDF can cause electron cyclotron maser instabilities and generate multiple harmonic electron cyclotron waves due to a wave-particle interaction at the relativistic resonance condition $\omega=n\Omega_{ce}/\gamma_d+u_{d\parallel}k_{\parallel}$. Here the Lorentz factor $\gamma_d$ is due to the motion of the electron beam, $u_{d\parallel}$ the parallel drift speed of the electron beam.
\end{itemize}

%%%%%%%%%%%%%%%%%%%%%%%%%%%%%%%%%%%%%%%%%%%%%%%%
% Results
%%%%%%%%%%%%%%%%%%%%%%%%%%%%%%%%%%%%%%%%%%%%%%%%
\section{Results}\label{sec:results}

We presented now the results of the simulations described above, which were initialized with either a two-streaming EVDF (see \myreffig{fig:EVDF_Yao}(a1,a2)) or a perpendicular crescent-shaped EVDF (see \myreffig{fig:EVDF_Yao}(b1,b2)).
Those initial EVDFs lead to the following instabilities, respectively,
\begin{itemize}
    \item Electron/ion streaming instabilities. They cause (electrostatic) ion-acoustic waves. In addition, the relative streaming also generates electrostatic Langmuir waves. Transverse emission in the form of harmonics of Langmuir waves is generated via a non-linear wave--wave interaction process.
    \item Electron cyclotron maser instabilities (ECMIs). They cause multiple harmonic electromagnetic ECWs due to a wave--particle process.
\end{itemize}

\subsection{Evolution of the instabilities}

Here we briefly describe the evolution of both instabilities to understand the evolution of the beam-plasma system and the resulting radiation properties from magnetic reconnection EVDFs.
More detailed stability analysis have been already carried out in the past. See the references in the Introduction and \citet[][]{Yao2021,Zhou2020}.
We focus, in particular, on the instabilities driven by the parallel drift speed, since they are the most relevant for the generation of electromagnetic escaping waves.

\begin{figure}[ht!]
  \plotone{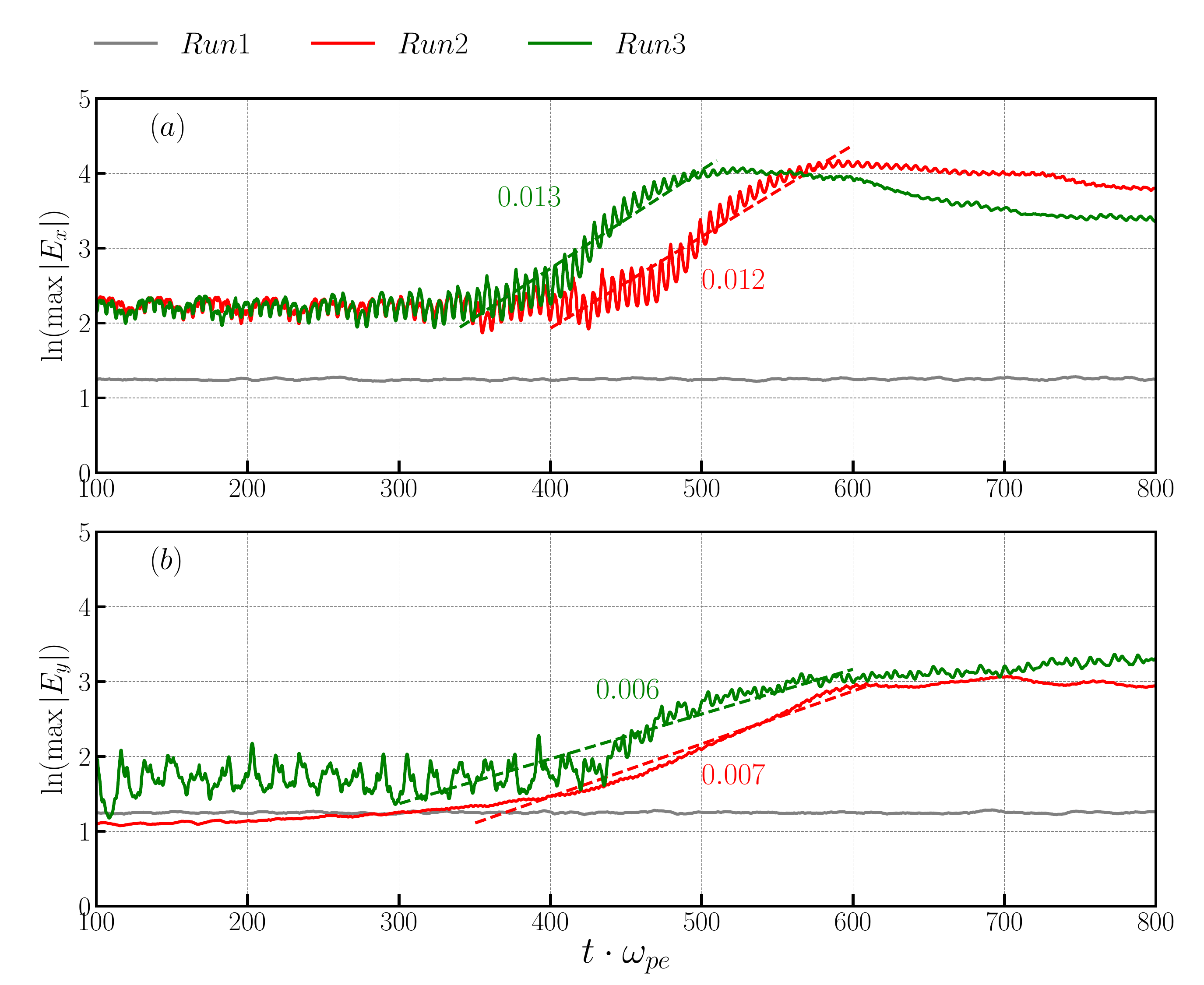}
  \caption{Temporal evolution of the maximum amplitude of the electric field oscillations (a) in $x$ direction $\ln|\max(E_x)|^2$ and (b) in $y$ direction $\ln|\max(E_y)|^2$ for all simulations. The dashed lines indicate fittings to the linear growth phase, the numbers next to them indicate the growth rates values (in units of $\omega_{pe}$).}
  \label{fig:growth_rate_3runs}
\end{figure}

As time evolves, the kinetic energy of the electron beam is transferred into the thermal energy of the beam-plasma system and electric field energy. Meanwhile, the variation of magnetic energy is negligible \citep[][]{Yao2021}.
This way the signatures of the unstable waves due to this instability can be seen mostly in the electric field energy. 
\myreffig{fig:growth_rate_3runs} (a) shows the temporal evolution of $\ln|max(E_x)|^2$ (the electric field component parallel to the background magnetic field) for three runs.
For the control case Run1 (with electron beam), the fluctuations of the electric field component $E_x$ always slightly oscillate in the thermal noise level.
The other two runs with a FEB, namely, Run2 and Run3, have initially larger electric field oscillations than those of Run1, and they are also similar to each other. That indicates that this enhanced initial level of electric field oscillations is only due to the parallel drift speed of the beam, which contributes to the relative streaming between the total EVDF and the ion VDF (initially at rest), as we explain later.

The Maxwellian electron beam of Run2 causes longitudinal electric field oscillations $\ln|max(E_x)|^2$ which exponentially grow between $t\approx 400\omega_{pe}^{-1}$ and $t\approx 600\omega_{pe}^{-1}$, with a estimated linear growth rate of $0.012\omega_{pe}$. This linear growth phase is followed by a saturation, where the electric field energy remains roughly constant. This implies that the source of free energy is depleted.

\begin{figure}[ht!]
  \plotone{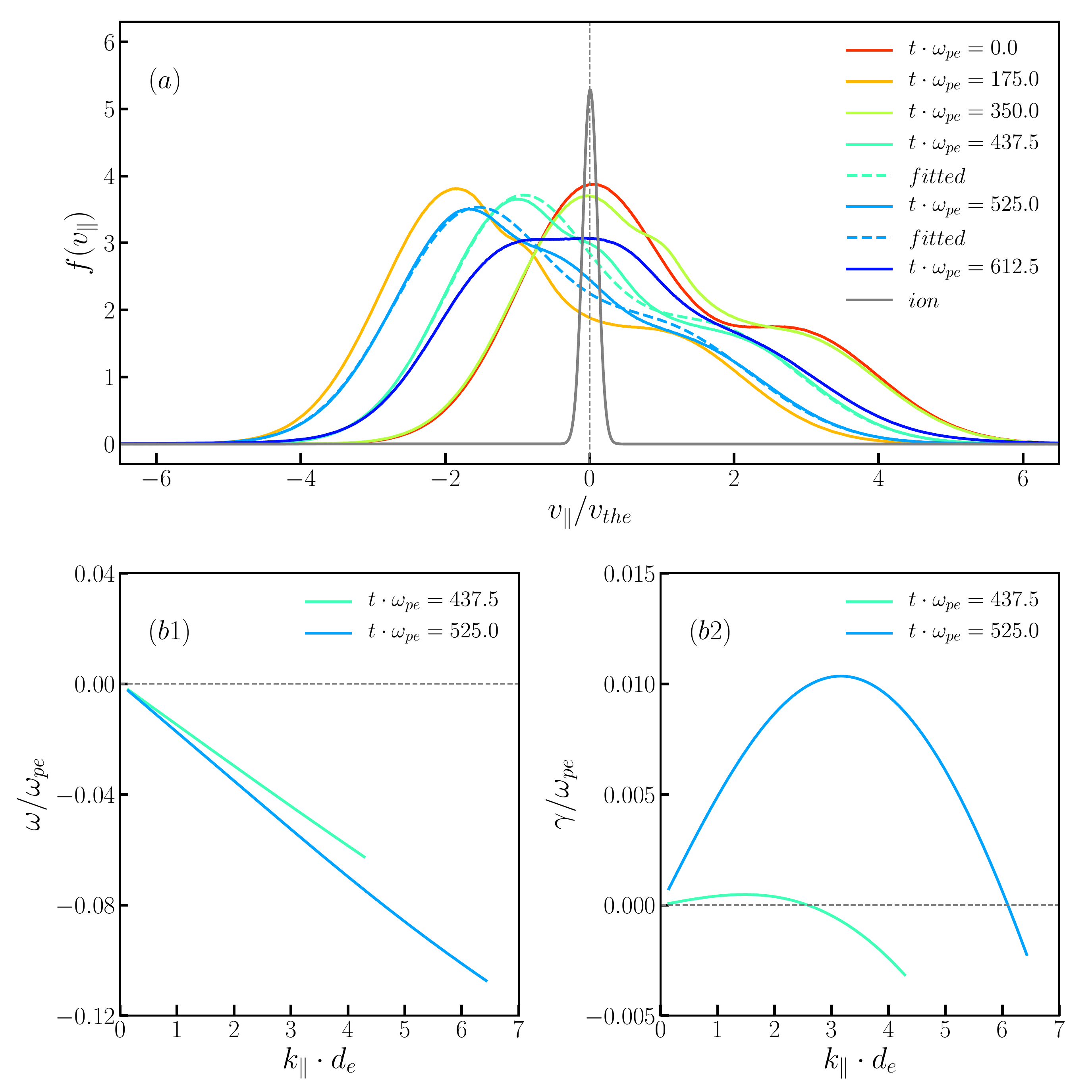}
  \caption{(a) 1D parallel total electron (sum of background and beam electrons) velocity distribution function (VDF) at different times for Run2. The ion VDF (solid grey curve) nearly does not vary as time evolves. Note the ion VDF is scaled by a factor of 1/40 to visualize all the VDFs in the same panel. 1D parallel electron VDF at $t= 437.5\omega_{pe}^{-1}$ and $t= 525.0\omega_{pe}^{-1}$ are fitted by \myrefeq{eq:model_evdf} (dashed curves). Unstable solutions of \myrefeq{eq:dispersion_relation} for the fitted VDFs at $t= 437.5\omega_{pe}^{-1}$ and $t= 525.0\omega_{pe}^{-1}$ are shown in the bottom panels: (b1) frequency vs wavenumber (i.e. $\omega-k_{\parallel}$) and  (b2) growth rate vs wavenumber (i.e. $\gamma-k_{\parallel}$).}
  \label{fig:VDFs_growth_rate}
\end{figure}

In order to investigate the specific source of free energy for this observed electric field growth in Run2, and so the instabilities themselves, it is necessary to analyze the distribution functions. 
\myreffig{fig:VDFs_growth_rate} (a) shows the time evolution of 1D EVDFs along the parallel direction (i.e. $x$ direction) for Run2.  The ion VDF (see the solid grey curve in \myreffig{fig:VDFs_growth_rate} (a)) is also present at $v_x=0$. We verified that the ion VDF practically does not vary during the entire evolution of the system.

The initial total EVDF (sum of background and beam electrons) at $t=0.0$ does not feature a positive velocity gradient and therefore is stable.
But it has a net drift or bulk flow speed $v_x>0$. That,  combined with the fact that the ion VDF is located at $v_x=0$, results in a net relative drift speed between the ion VDF and EVDF, equal to $V_{ei}=0.95v_{the}$.
This relative drift represents a net current and is a possible source of free energy for instabilities and waves.
Due to Ampere's law, this causes the finite initial electric field amplitude for Run2 in \myreffig{fig:growth_rate_3runs}, which represents strong electric field oscillations accompanied by associated oscillatory motions of the EVDF. The initial EVDF only slightly changes during the time interval
in which the electric field energy remains constant in \myreffig{fig:growth_rate_3runs}, i.e., up to $t\approx 400\omega_{pe}^{-1}$.
The electric field oscillations occur at the plasma frequency modified by thermal effects, which can be identified as Langmuir waves.

After $t\approx 400\omega_{pe}^{-1}$ the EVDF shifts to the left, i.e., the bulk flow speed of the whole EVDF decreases and adopts negative values in order to decrease the net current (relative electron-ion streaming).
As a consequence, the relative drift speed between electrons and ions becomes negative as well, reaching values up to $V_{ei}=-0.71v_{the}$ at $t\approx 525.0\omega_{pe}^{-1}$.
This makes the electron-ion distribution unstable (see discussion below), since the bulk of the ion VDF falls into the monotonously decreasing part (or to the right side) of the EVDF.
This unstable distribution is correlated with the exponential growth phase denoted by the red dashed line of \myreffig{fig:growth_rate_3runs}.
As a consequence the source of free energy, the relative electron-ion drift speed, is exhausted quickly, so that the final EVDF shown in \myreffig{fig:VDFs_growth_rate}(a) for $t\approx 612.5\omega_{pe}^{-1}$ moves to a location closer to the initial EVDF. This final EVDF also features a larger temperature (broader distribution).

In order to numerically quantify the growth rates of the instability driven by the relative electron-ion drift, we linearize the 1D electrostatic  Vlasov equation for three plasma species, ion VDF (subscript $i$), background EVDF (subscript $ec$) and beam EVDF (subscript $eb$). We assume for each one a 1D drifting Maxwellian in the following way
\begin{align}\label{eq:model_ivdf}
f_i(v_x) &= \frac{n_i}{\sqrt{2\pi}v_{thi}}\exp\left(\frac{(v_x -u_{d,i})^2}{2v_{thi}^2} \right),\\
f_e(v_x) &= \frac{n_{ec}}{\sqrt{2\pi}v_{th,ec}}\exp\left(\frac{(v_x -u_{d,ec})^2}{2v_{th,ec}^2} \right) + \frac{n_{eb}}{\sqrt{2\pi}v_{th,eb}}\exp\left(\frac{(v_x -u_{d,eb})^2}{2v_{th,eb}^2} \right),\label{eq:model_evdf}
\end{align}
for ions and the total electron population, respectively.
The parameters are obtained from the fitting of the actual simulation distribution functions at different times.
The resulting dispersion relation yields \citep{Gary1993,Che2010}:
\begin{align}
    1+\sum_s\frac{\omega_{ps}^2}{k^2v_{th,s}^2}\left[1+\zeta_s Z(\zeta_s)\right]=0
    \label{eq:dispersion_relation}
\end{align}
where $s=i$, $ec$ or $eb$. $Z(\zeta_s)$ is the plasma dispersion function,
\begin{equation}
    Z(\zeta_s)=\frac{1}{\sqrt{\pi}}\int_{-\infty}^{\infty}\frac{\exp(-x^2)}{x-\zeta_s}dx,\qquad\mathfrak{Im}(\zeta_s)>0
\end{equation}
with arguments:
\begin{align}
    \zeta_s&=\frac{\omega-ku_{d,s}}{\sqrt{2}kv_{th,s}}
\end{align}
Here $\omega_{ps}$ is the plasma frequency, $v_{th,s}$ the thermal speed and $u_{d,s}$ the drift speed of the $s$-th species, respectively.
In this subsection we normalize our results to the electron plasma frequency $\omega_{pe}$, electron thermal speed $v_{the}=\sqrt{k_BT_e/m_e}$ and electron number density $n_e$ of the initial ($t=0$) electron background plasma (otherwise indicated with the subscript $ec$).
% The initial electron Debye length of the electron background is thus $\lambda_{De}=v_{the}/\omega_{pe}$.

Note that we solve \myrefeq{eq:dispersion_relation} for a complex $\omega+i\gamma$, where $\omega$ is the real frequency and $\gamma$ the growth rate (or damping rate) and for a given $k$.
We search for the complex roots ($\omega+i\gamma$,$k$) that satisfy the dispersion relation \myrefeq{eq:dispersion_relation} by means of a multidimensional root finding algorithm.
This method can find all infinite roots of \myrefeq{eq:dispersion_relation},
which are mostly heavily damped.
We are only interested in the solutions with positive $\gamma$, namely, growing waves.

For this purpose, we first fitted the simulation VDFs shown in \myreffig{fig:VDFs_growth_rate}
by the ion VDF \myrefeq{eq:model_ivdf} and the EVDF \myrefeq{eq:model_evdf}.
Note that the ion VDF does not practically vary during the evolution of the system.
The solutions of  \myrefeq{eq:dispersion_relation}  show that most of the EVDFs shown in \myreffig{fig:VDFs_growth_rate} lead to stable solutions, i.e., negative or zero $\gamma$.
Only two (fitted) EVDFs, namely, at $t= 437.5\omega_{pe}^{-1}$ and $t= 525.0\omega_{pe}^{-1}$ (see dashed curves shown in \myreffig{fig:VDFs_growth_rate} (a)), lead to a positive growth rate $\gamma$.
% For EVDF at $t=437.5\omega_{pe}^{-1}$, it leads to to weak growth rates, i.g., two roots $\gamma_1$ and $\gamma_2$ (denoted by light-green solid/dashed curve in \myreffig{fig:VDFs_growth_rate}(b2)) are very close to zero in a magnitude $\le 5\times10^{-4}\omega_{pe}$.
The EVDF at $t=525.0\omega_{pe}^{-1}$ features a larger source of free energy, namely, a larger electron-to-ion drift speed and so a much larger growth rate than for $t= 437.5\omega_{pe}^{-1}$. The fitting parameters for this EVDF at $t=525.0\omega_{pe}^{-1}$ are drift speeds $u_{d,ec}=-1.87v_{the}$, $u_{d,eb}=0.4v_{the}$, electron beam-to-background density ratio $n_{eb}/n_{ec}=1.04$, thermal speeds $v_{th,ec}=1.007v_{the}$ and $v_{th,eb}=1.573v_{the}$.
The solution of \myrefeq{eq:dispersion_relation} with those parameters
(see skyblue solid curve in \myreffig{fig:VDFs_growth_rate}(b2)) shows a positive growth rate $\gamma>0$ in the wavenumber range about $k_{\parallel}\in[0,6d_e^{-1}]$ with a maximum $\gamma\approx 0.01\omega_{pe}$, which is in rough agreement with the value $0.012\omega_{pe}$ obtained from the linear fitting of the electric field amplitude in \myreffig{fig:growth_rate_3runs} (a).
The difference can be attributed to the dynamical evolution of the EVDF during the linear growth phase; namely, it is hard to choose a representative set of parameters  to solve \myrefeq{eq:dispersion_relation} for the linear theory analysis within this time period.

\myreffig{fig:VDFs_growth_rate} (b1) shows the real frequency vs the wavenumber of the modes shown in \myreffig{fig:VDFs_growth_rate} (b2). The unstable branches at $t=437.5\omega_{pe}^{-1}$ (solid light-green curve) and $t=525.0\omega_{pe}^{-1}$ (solid skyblue curve) belong to the same wave mode with a mostly linear dependence between frequency and wavenumber. Its phase speed (or slope) is negative and equal to  $\omega/k_{\parallel}\approx-0.25v_{the}=-0.0175c$ (at $t=525.0\omega_{pe}^{-1}$).
%Resonant interaction
This implies the existence of a resonant interaction between the left tail of the ion VDF and the EVDF at such a phase speed. The source of free energy comes from the positive velocity gradient of the ion VDF combined with the negative velocity gradient of the EVDF.
%
% For the first branch $\omega_1$ for the fitted EVDF   and the branch $\omega$ at $t=525.0\omega_{pe}^{-1}$, they are mostly linear dependent  on $k_{\parallel}$ with a negative phase speed (or slope) equal to
%Ion acoustic waves
The ion acoustic speed, defined as \citep{Baumjohann1997,Swanson2003a}:
\begin{equation}\label{eq:iaw_speed}
c_{s}^2 =\frac{k_BT_e+3k_BT_i}{m_i},
\end{equation}
is $c_{s} \approx\pm0.014c$, which  in roughly agreement with the theoretical phase speed of the unstable wave mode.
The differences can be attributed to the non-Maxwellian nature of the EVDF, which in practice leads to a larger effective electron temperature and thus increases the effective sound speed. Therefore, we can identify the unstable wave mode as ion-acoustic waves (see Section \ref{sec:Langmuir_waves}): strong ion density fluctuations that start to develop during the unstable period after $t\approx 400\omega_{pe}^{-1}$, with frequency and wavenumber matching the linear dispersion relation of ion acoustic waves. Note that ion acoustic waves are the essential ingredient, besides Langmuir waves, to generate the fundamental plasma emission via wave-wave interactions (see Section \ref{sec:introduction}).

Those agreements between linear theory calculations and simulation results are evidence that supports the development of an electron/ion instability with the ion-acoustic branch as the unstable mode in our system.
This is despite the fact that our system does not satisfy the standard threshold for such an instability, which usually requires $T_e\gg T_i$ and it is based on a single drifting Maxwellian for the whole population \citep{Gary1993,Treumann2001}.
For a plasma with equal electron and ion temperatures, like the initial conditions of our simulations, ion-acoustic waves are actually heavily Landau damped, so they are not observed in the early time period (see \myreffig{fig:IA_waves}(a)).
It is only later in the evolution of the system (after $t\approx 400\omega_{pe}^{-1}$) that the strong non-Maxwellian EVDF, which is hotter than the initial one, with a heavy skewed tail and interacts with the ion EVDF, can allow the excitation of ion-acoustic waves with an amplitude that exponentially grows.

The linear growth phase of the longitudinal $E_x$ component of Run2 is also associated with an exponential growth of the transverse component $E_y$. This is the component responsible for the electromagnetic radiation, which is observed as harmonics of the electron plasma frequency in the electromagnetic wave branch (see Figures \ref{fig:EM_PE} and \ref{fig:escaping_waves}).
That means the excitation of ion-acoustic waves during the linear growth phase of $E_x$ is correlated with the excitation of transverse electromagnetic waves (harmonics) as seen in $E_y$.
Note, however, that the estimated linear growth rate is $0.07\omega_{pe}$ and therefore smaller than for $E_x$, indicating that those waves are only indirectly caused by the relative electron-ion drift speed.

Run3 with a crescent-shaped EVDF has a somewhat different evolution.
The evolution of the longitudinal electric field component $E_x$ also has a linear growth phase, similar to Run2. It starts, however, a bit earlier $t\approx 300\omega_{pe}^{-1}$ and ends near  $t\approx 400\omega_{pe}^{-1}$ with the  same saturation level as that of Run2 and with a similar growth rate as well: $0.013\omega_{pe}$.
That implies that the linear growth phase is also driven by the relative electron-ion streaming, since the parallel EVDFs for both Run2 and Run3 have the same parallel drift speed.
The earlier start of the linear phase for the evolution of the $E_x$ component of Run3 can be attributed to the perpendicular EVDF, which is different from Run2.
Indeed, the transverse component $E_y$ for Run3 grows from the very beginning, saturating at the same time ($t\approx 400\omega_{pe}^{-1}$) as the longitudinal component.
This implies that the positive perpendicular velocity gradients generate unstable waves from the very beginning due to the electron cyclotron maser instability, as shown below.

%%%%%%%%%%%%%%%%%%%%%%%%%%%%%%%%%%%%%%%%%%%%%%%%
% Langmuir waves
%%%%%%%%%%%%%%%%%%%%%%%%%%%%%%%%%%%%%%%%%%%%%%%%
\subsection{Langmuir waves}\label{sec:Langmuir_waves}

The introduction of beam plasma to the background plasma has a significant influence on the local plasma frequency for Run2 and Run3, because the local electron number density is heavily affected by the density fluctuations as the electron beam propagates through the ambient plasma. By integrating the power spectral density (PSD) of $E_x$ over the wavenumber $k_{\parallel}$, yields the power spectra with respect to frequency, namely
\begin{equation}
  \mathcal{P}(\omega)=\int |E_x\left(k_{\parallel},\omega\right)/B_0|^2 dk_{\parallel}
\end{equation}
The local plasma frequency $\omega_{loc}$ can then be numerically determined by the corresponding frequencies of those maxima of the $\mathcal{P}(\omega)$ \citep[see more details in][]{Yao2021}. In this study, the effective local plasma frequency is about $\omega_{loc}\approx 0.95\ \omega_{pe}$ both for Run2 and Run3, while $\omega_{loc}=\omega_{pe}$ for Run1.

Harmonics of Langmuir waves at the frequency of multiples of the local plasma frequency, i.e., $\omega=n\omega_{loc}\ (n=1,2,3,...)$, due to two-streaming EVDFs are observed both in Run2 and Run3. In this part, we analyze the results of Run2 as an example to explain the generation of Langmuir waves. Similar phenomena are also observed in Run3.

\begin{figure}[ht!]
  \plotone{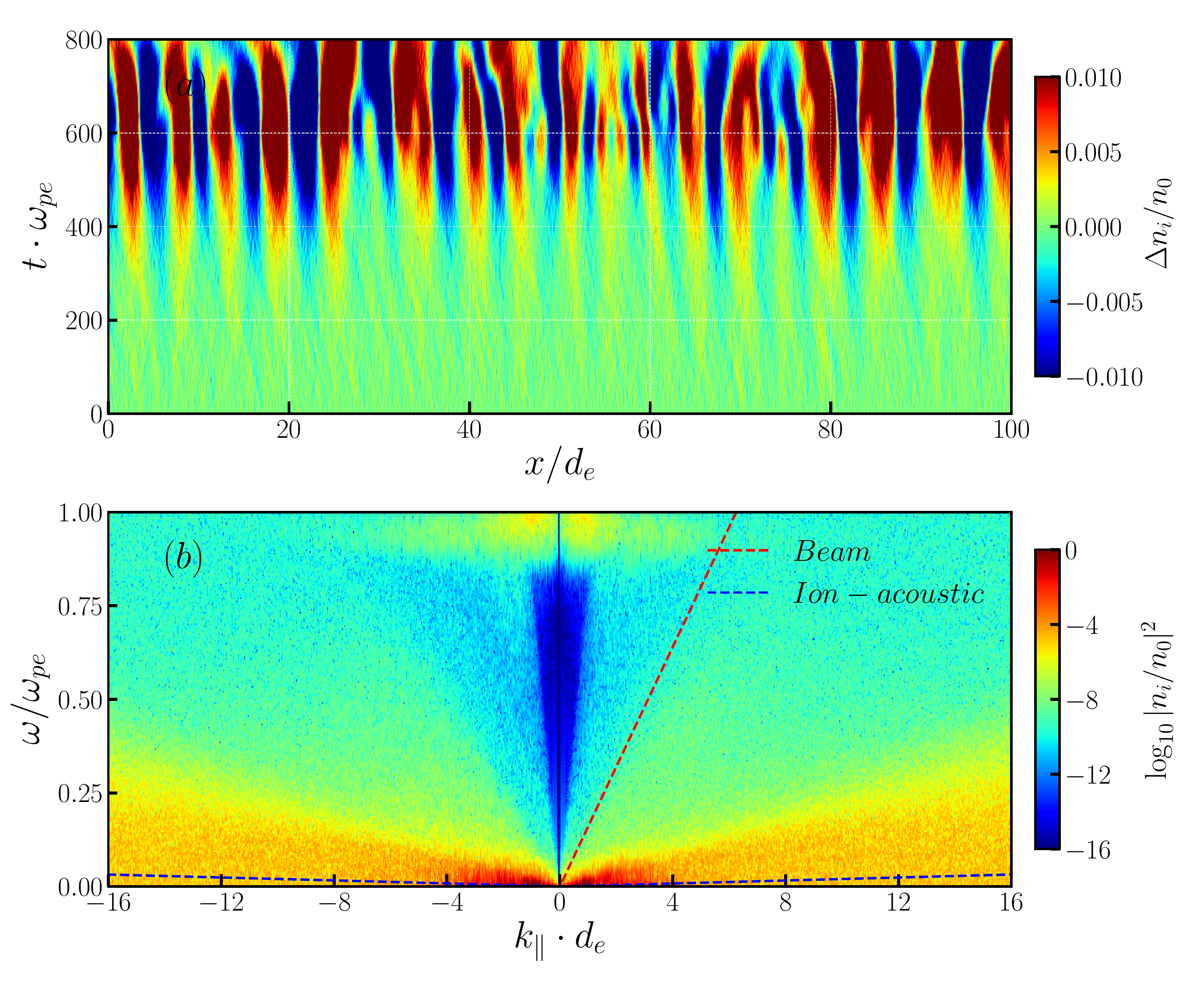}
  \caption{(a) Ion number density fluctuation $\Delta n_i=n_i-\overline{n_i}$ in $x-t$ plane during time $t\cdot \omega_{pe}=0-800$. The ion density fluctuation $\Delta n_i$ is normalized by $n_0$. (b) Corresponding PSD of the normalized ion number density $n_i(k_{\parallel},\omega)/n_0$. Here the PSD is estimated by $\log_{10}|n_i/n_0|^2$. The beam mode (red dashed line) and the ion-acoustic mode (blue dashed line) are overlaid. All data are from simulation Run2. \label{fig:IA_waves}}
\end{figure}

Due to the small mass ratio $\mu=m_i/m_e=100$ used in this study, ion-acoustic waves can be generated within the time scale of our simulations.
They start to be visible from $t\cdot\omega_{pe}\approx 100$ on (see \myreffig{fig:IA_waves}).
\myreffig{fig:IA_waves} (a) shows the ion number density fluctuation $\Delta n_i=n_i-\overline{n_i}$ within the time interval $t\cdot \omega_{pe}=0-800$ in the $x-t$ plane. 
\myreffig{fig:IA_waves} (b) shows the power spectral density (PSD) derived from the ion number density $n_i(k_{\parallel},\omega)$ in Fourier space.
The PSD of $n_i(k_{\parallel},\omega)$ in the low-frequency region is broadly thermally spread near the dispersion relation curve of ion-acoustic waves \citep[][]{Baumjohann1997, Swanson2003a}:
\begin{equation}
  \omega^2=\frac{k_BT_e}{m_i}\frac{k^2}{1+k^2\lambda_D^2} + 3\frac{k_BT_i}{m_i}k^2=\frac{c_s^2k^2}{1+3T_i/T_e}\left(\frac{1}{1+k^2\lambda_{D}^2}+3\frac{T_i}{T_e}\right)\label{eq:disp_rela_S}
\end{equation}
here $c_s$ is the ion-sound speed defined in \myrefeq{eq:iaw_speed}. Note that this expression contains short-wavelength corrections (contained in the denominator $1+k^2\lambda_D^2$). In the long-wavelength limit ($k\lambda_D\ll1$, \textbf{or $kd_e\ll14.28$}) this expression reduces to $\omega=c_s k$. All calculations and results in this work  satisfy this approximation very well\textbf{, since the spectral power is confined to $k\cdot d_{e}<4$ as seen in \myreffig{fig:IA_waves}(b)}.

This way, spectral power near this branch implies the existence of low-frequency ion-acoustic waves $S$.

\begin{figure}[ht!]
  \plotone{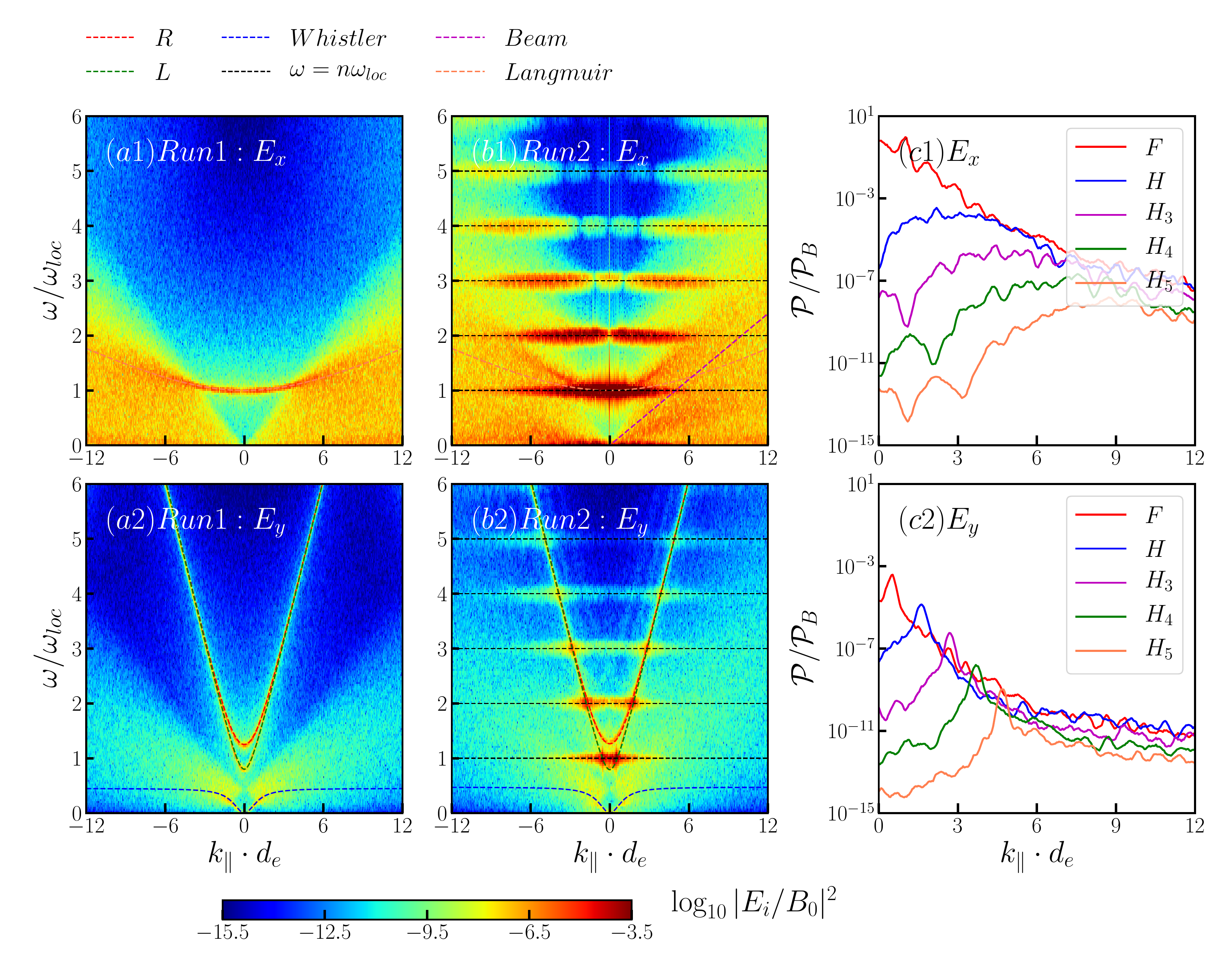}
  \caption{Power spectral density (PSD) for Run1 (a1--a2) and Run2 (b1--b2) in the time window $t\cdot \omega_{pe}=455-634$. PSDs estimated by $\log_{10}|E_i/B_0|^2\ (i=x,y)$ are displayed in the $k_{\parallel}-\omega$ plane. Dispersion relation curves of Langmuir mode (orange dashed curve), harmonics of the plasma frequency $\omega=n\omega_{loc}$ (black dashed lines), beam mode (magenta dashed curve), R (red dashed curve), L (green dashed curve) and Whistler (blue dashed curve) are overlaid. Integrated power spectra $\mathcal{P}(k_{\parallel})$ of corresponding harmonics (fundamental $F$, harmonic $H$ and higher harmonics $H_n\ (n=3,4,5)$) of Langmuir waves in (c1) $E_x$ and (c2) $E_y$ for Run2 are shown in $k_{\parallel}-\mathcal{P}$ plane. The power spectrum $\mathcal{P}$ is normalized by the initial magnetic power density $\mathcal{P}_B=|B_0|^2$. \label{fig:EM_PE}}
\end{figure}

\myreffig{fig:EM_PE} shows a comparison of power spectral densities between Run1 and Run2. For Run1, which works as our control case for a thermal plasma without an energetic electron beam, we only see the electrostatic Langmuir mode in the $E_x$ component, while the R, L, and whistler modes in the $E_y$ component.
The electrostatic Langmuir waves are fitted by the standard Bohm-Gross dispersion relation, i.e., $\omega=\sqrt{\omega_{pe}^2+3v_{the}^2k^2}$ \citep[][]{Bohm1949}.
The dispersion relations of R, L, and whistler modes are solved by using the dispersion relation of waves in the cold plasma approximation \citep[][]{Stix1992,Swanson2003a}.

For Run2, we found harmonics of electrostatic Langmuir waves in the parallel direction, i.e., $k_{\parallel}$, (\myreffig{fig:EM_PE}(b1)) and harmonics of the transverse waves in the perpendicular direction i.e., $k_{\perp}$, (\myreffig{fig:EM_PE}(b2)).
In \myreffig{fig:EM_PE}(b2), the dispersion relation curves of R, L and Whistler modes are also overlaid. The dispersion relation curves of the electromagnetic R and L modes approach and merge when $\omega\ge 2\omega_{loc}$. The multiple harmonic wave modes with a parallel phase speed equal to or faster than the speed of light in the long-wavelength regime, i.e., the superluminal regime $\omega\ge ck_{\parallel}$, are the transverse waves.

By assuming that the power spectrum of a wave mode follows a Gaussian power distribution along its dispersion relation curve in the $k_{\parallel}-\omega$ domain, the integrated power distribution for each harmonic of Langmuir waves can be quantitatively estimated as follows:
\begin{equation}
    \mathcal{P}_i\left(k\right)=\int \frac{1}{\sqrt{2\pi \sigma}}\exp\left[-\frac{\left(\omega(k)-\omega_i\right)^2}{2\sigma^2}\right]|E_i(k,\omega)|^2d\omega
    \label{eq:PSD_RS}
\end{equation}
here $\sigma=0.02\omega_{pe}$ is the frequency width spread along the theoretical dispersion curve.

\myreffig{fig:EM_PE} (c1) shows this integrated power spectra $\mathcal{P}_i\left(k\right)$ versus wavenumber $k_{\parallel}$ of harmonics of the electrostatic Langmuir waves for Run2.
The power spectrum of the fundamental Langmuir waves decrease from a magnitude of $10^{-1}$ at $k_{\parallel}=0$ to $10^{-7}$ at $k_{\parallel}\cdot d_e=12$. But most of the power is concentrated between $kd_e=0$ and $kd_e=0.96$, both of which represent a local maxima.
Meanwhile, the power spectra of the second to fifth harmonic of electrostatic Langmuir waves increase to their maximum and then decrease to a magnitude of $10^{-7}$ at $k_{\parallel}\cdot d_e=12$.
As the harmonic number increases, the magnitude of power spectrum of each harmonic electrostatic Langmuir waves decreases significantly, roughly by two orders of magnitude.
Similar phenomena are found in transverse emission at frequency of multiple $\omega_{loc}$.
\myreffig{fig:EM_PE} (c2) shows the power spectra $\mathcal{P}(k_{\parallel})$ of harmonics of transverse waves at the frequency of multiple local plasma frequencies.
They increase to their maximum and then decrease to a magnitude of $10^{-11}$ at $k_{\parallel}\cdot d_e=12$.
For fundamental waves, a ``hump'' in the power distribution curve corresponds to a wave interaction with the \textbf{$L$} mode at $k_{\parallel}\cdot d_e\approx \textbf{$0.48$}$ (see the red curve in \myreffig{fig:EM_PE} (c2)). In general, the humps represent an enhancement of the spectral power at the intersection of wave modes indicating a wave--wave interaction process \textbf{(see more details at the end of Section~\ref{sec:transverse})}.
While for the higher harmonic transverse waves, humps form at $k_{\parallel}\cdot d_e\approx \textbf{$1.67$, $2.67$, $3.68$, $4.69$}$ respectively. They are related to a wave interaction with the R mode. This implies the wave interactions with L and R modes have a significant influence on the power spectra of harmonics of transverse waves.

\begin{figure}[ht!]
  \plotone{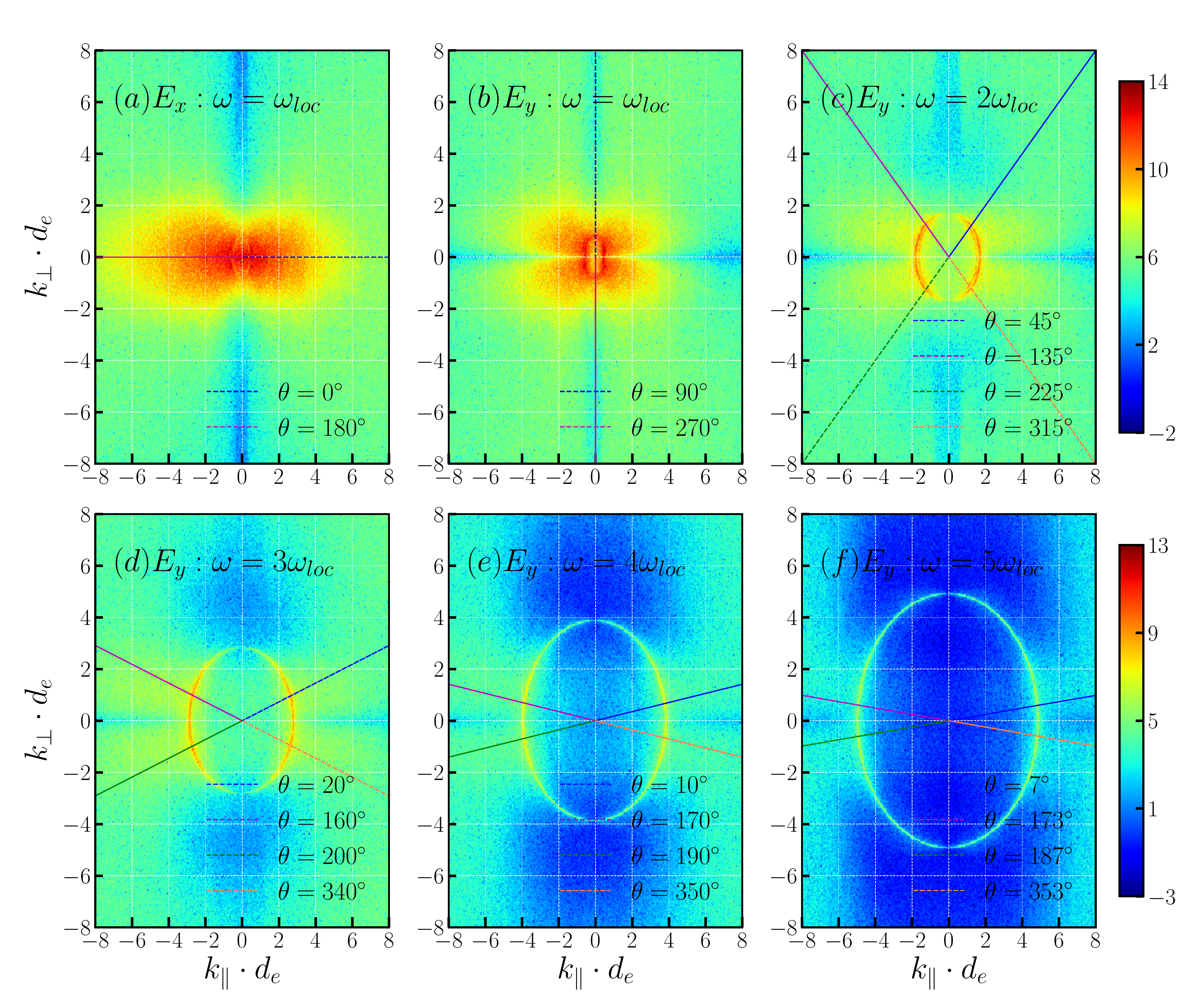}
  \caption{PSD in $k_{\parallel}-k_{\perp}$ plane of (a) electrostatic Langmuir waves at $\omega=\omega_{loc}$ and (b--f) fundamental ($\omega=\omega_{loc}$) to fifth harmonic ($\omega=5\omega_{loc}$) of transverse waves, respectively. The radiation intensity is estimated by $\log_{10}|E_i(\omega,k_{\parallel},k_{\perp})/B_0|^2\ (i=x,y)$ at given $\omega=n\omega_{loc}$ for Run2 in the time window $t\cdot \omega_{pe}=455-831$, with $n=1,2,...,5$ the harmonic number. Here $\theta=\arctan \left(k_{\perp}/k_{\parallel}\right)$ is the wave propagation angle. These dashed oblique lines indicate the symmetry axis of the angle region where the maximum radiation intensity of electrostatic Langmuir waves and transverse waves take place (see Appendix ~\ref{app:Radiation_pattern}).\label{fig:EM_PE_3D}}
\end{figure}

\myreffig{fig:EM_PE_3D} shows the PSD of $E_x$ for electrostatic Langmuir waves and of $E_y$ for the fundamental and up to the fifth harmonic of transverse waves in the $k_{\parallel}-k_{\perp}$ plane.
The PSD of the electrostatic Langmuir waves and these transverse harmonics are anisotropic, and their radiation intensities are significantly angle dependent.
An analytical analysis of radiation intensity pattern due to the most probable wave--wave interactions for multiple harmonic transverse emission is discussed in Appendix ~\ref{app:Radiation_pattern}. We solved those equations for the radiation intensity in dependence on the angle. The results are summarized in \myreftab{tab:radiation_angle}.
The dashed oblique lines in each panel of \myreffig{fig:EM_PE_3D} indicate the symmetry axis of the angle range where the maximum radiation intensity is theoretically expected to take place.
For example, the radiation intensity of the electrostatic Langmuir waves $L$ and backward scattered Langmuir waves $L'$ (propagating in the negative $k_{\parallel}$ direction) are mainly distributed in a range about the axis of symmetry $\theta=0^{\circ}$ in the right half-plane $k_{\parallel}>0$ and $\theta=180^{\circ}$ in the left half-plane $k_{\parallel}<0$ with a broad angle width spread (\myreffig{fig:EM_PE_3D}(a)).

For the fundamental transverse mode, the maximum radiation intensity takes place about the angle $\theta=90^{\circ}$ in the top half-plane and $\theta=270^{\circ}$ in the bottom half-plane with broad width spread in angle (see \myreffig{fig:EM_PE_3D}(b)).
For the harmonic transverse emission $H$ (see \myreffig{fig:EM_PE_3D}(c)), the maximum wave intensity mainly occurs at each quadrant in the $k_{\parallel}-k_{\perp}$ plane, e.g., around angles $\theta=45^{\circ},135^{\circ},225^{\circ},315^{\circ}$ respectively.
The wave radiation intensity of the third transverse emission $H_3$ (see \myreffig{fig:EM_PE_3D}(d)) are observed in an angle range about angles $\theta=20^{\circ},160^{\circ},200^{\circ},340^{\circ}$ and with a broad spread width in angle about $\Delta \theta \sim 15^{\circ}$.
For the fourth transverse emission (see \myreffig{fig:EM_PE_3D}(e)), the wave radiation intensity mainly occurs in the ranges $\theta=10^{\circ}\pm 5^{\circ},170^{\circ}\pm 5^{\circ},190^{\circ}\pm 5^{\circ},350^{\circ}\pm 5^{\circ}$ respectively.
While for the fifth transverse emission, its radiation intensity is weak in the transverse direction comparing to that of lower harmonic transverse emission, and it mainly appears in the angle range $\theta=7^{\circ}\pm 3^{\circ},173^{\circ}\pm 3^{\circ},187^{\circ}\pm 3^{\circ},353^{\circ}\pm 3^{\circ}$, respectively.

\begin{figure}[ht!]
  \plotone{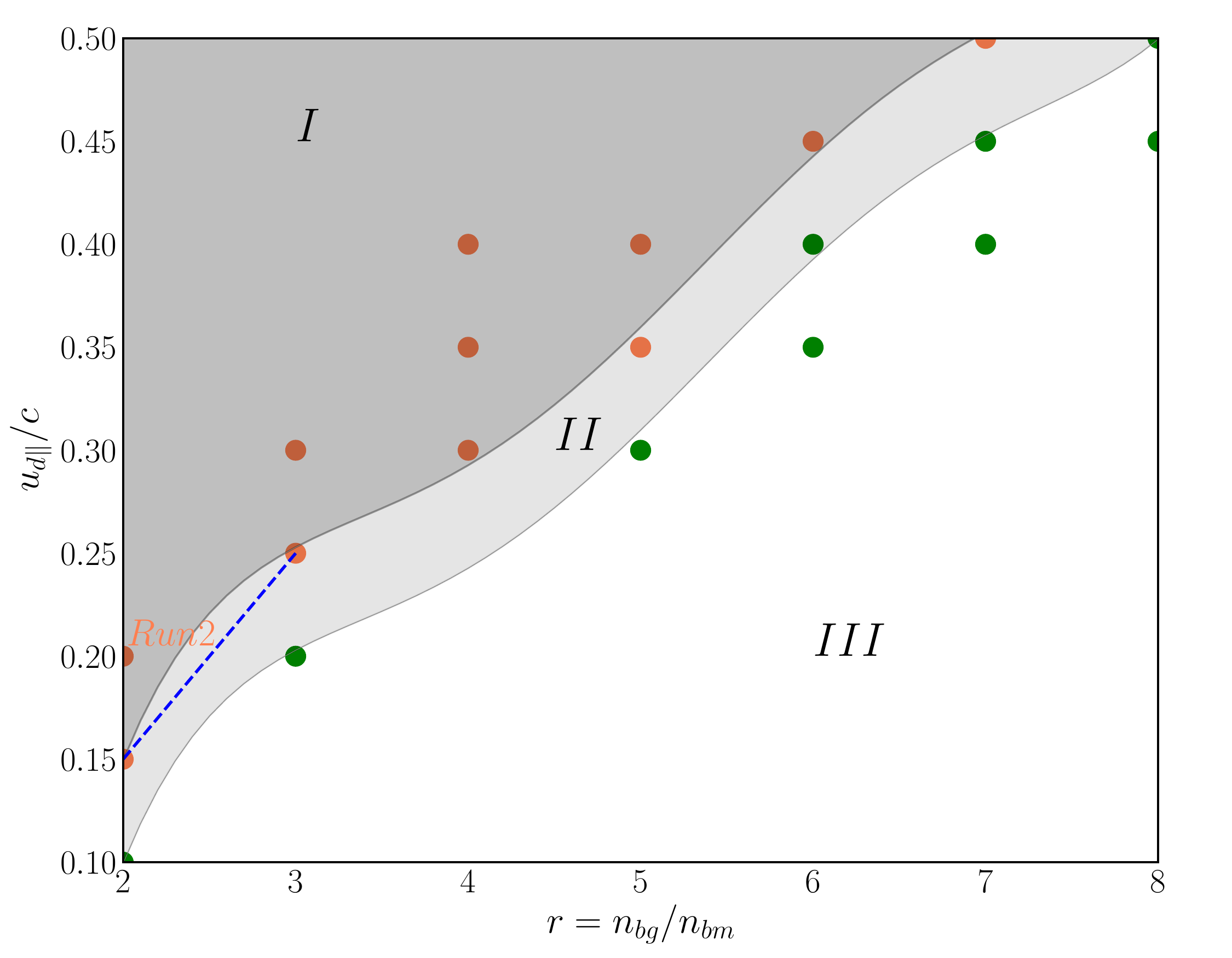}
  \caption{Parametric regime of ratio of background-to-beam plasma particle number density $r=n_{bg}/n_{bm}$ and parallel drift speed of the electron beam $u_{d\parallel}$ for the generation of harmonic plasma emission. Regime I: Harmonics of plasma emission can be generated, Regime III: transverse emission is not generated or at most the fundamental emission can be generated, Regime II: a transition region. Each point (denoted by green/orange dots) corresponds to the parameter pair ($r,u_{d\parallel}$) for a simulation. Orange dots indicate where harmonics of plasma emission are generated, while green dots imply where transverse emission is not generated or at most the fundamental emission can be generated. Blue dashed line: $u_{d\parallel}/c\ge 0.1n_{bg}/n_{bm}-0.05$ (see explanation in the text). Other parameters are same to  those of Run2.\label{fig:parametric_regime_plasma_emission}}
\end{figure}

We found that the excitation of multiple harmonic plasma emission sensitively depends on the background-to-beam plasma particle number density ratio $r=n_{bg}/n_{bm}$ (or its inverse $n_{bm}/n_{bg}$) and the field-aligned drift speed of the non-thermal beam $u_{d\parallel}$. We thus carried out numerical experiments to test the dependence of the excitation of plasma emission on the background-to-beam number density ratio $r= 2-8$ and the parallel drift speed of the non-thermal beam $u_{d\parallel}= 0.1-0.5c$.
Note that the parallel drift speed of the non-thermal electron beams generated in the antiparallel magnetic reconnection simulation presented above is within the range $u_{d\parallel}= [-0.2c, 0.2c]$, while the background-to-beam plasma particle number density ratio can be up to $r=n_{bg}/n_{bm}\ge 2$.
The observations show that the drift speed of the non-thermal electron beam is possibly between $[0.1c,0.5c]$ \citep[][]{Wild1959,Alvarez1973}.
\myreffig{fig:parametric_regime_plasma_emission} shows an empirical parametric regime of parameter pairs $(r,u_{d\parallel})$ based on 20 additional simulations similar to Run2.
Due to limitations in computational resources, those runs are implemented at several discrete parameter pairs $(r,u_{d\parallel})$ (denoted by orange and green dots in \myreffig{fig:parametric_regime_plasma_emission}), while other parameters are the same as those of Run2.
Based on the formation of harmonics of transverse Langmuir waves, we separate the $r-u_{d\parallel}$ plane into three regimes: Regime I -- where harmonics of transverse emission can be generated, Regime III -- where transverse emission is not generated or at most the fundamental emission can be generated, as well as a transition Regime II.

As mentioned above, two-streaming EVDFs with $|u_{d\parallel}|\le 0.2c$ can be found in the diffusion region and separatrices. In these regions, the background-to-beam particle density ratio is usually $n_{bg}/n_{bm}\ge 2$. \myreffig{fig:parametric_regime_plasma_emission} implies that only when the linear relation $u_{d\parallel}/c\ge 0.1n_{bg}/n_{bm}-0.05$ (denoted by the blue dashed line) is roughly satisfied, two-streaming EVDFs generated in  magnetic reconnection can cause  electron/ion streaming instabilities
causing ion acoustic and Langmuir waves and, in this way, generate at least the harmonic plasma emission.

An analytical estimation of the threshold $(r,u_{d\parallel})$ for the generation of plasma emission, i.e., harmonics of electrostatic/transverse Langmuir waves, can in principle be constructed based on the understanding of energy conversion in the mode conversion process of plasma emission, which is beyond the scope of this study.

%%%%%%%%%%%%%%%%%%%%%%%%%%%%%%%%%%%%%%%%%%%%%%%%
% ECWs
%%%%%%%%%%%%%%%%%%%%%%%%%%%%%%%%%%%%%%%%%%%%%%%%
\subsection{Electron cyclotron waves}

In the perpendicular direction $k_{\perp}$, electrostatic Bernstein waves (EBWs) and ECMI-generated electromagnetic electron cyclotron waves (ECWs) co-exit.

The dispersion relations of Bernstein modes are solved from the following equation \citep[][]{Bernstein1958}:
\begin{equation}
  1-\frac{2\omega_{loc}^2}{\lambda_e}e^{-\lambda_e}\sum\limits_{n=1}^{\infty}\frac{n^2I_n(\lambda_e)}{\omega^2-n^2\Omega_{ce}^2}=0
  \label{eq:ElectronBernsteinMode}
\end{equation}
where $I_n(\lambda)$ is the modified Bessel function of the first kind with argument $\lambda_e=k_{\perp}^2v_{the}^2/\Omega_{ce}^2$, $v_{the}$ is the electron thermal speed of the background plasma. In the short wavelength regime, the frequency of the nth harmonic of Bernstein mode asymptotically approaches to $\omega\to n\Omega_{ce}$.

The dispersion relation of ECWs due to electron gyroresonance conditions is $\omega=n\Omega_{ce}$, this kind of gyroresonance-related electromagnetic waves are named as electron gyroresonance emission if their frequency exceeds the local plasma frequency of the ambient plasma and thus can escape out \citep[][]{Aschwanden2005}.

Electron beam with non-thermal EVDFs offering perpendicular sources of free energy can drive ECMIs, and generate electromagnetic ECWs at relativistic electron gyrofrequency and its harmonics \citep[][]{Melrose1986}, namely,
\begin{equation}
  \omega=\frac{n\Omega_{ce}}{\gamma_{d}}+u_{d\parallel}k_{\parallel}
\end{equation}
where the Lorentz factor $\gamma_{d}=\sqrt{1+\left(u_{d\parallel}^2+u_{d\perp}^2\right)/c^2}$ with $u_{d\parallel}$ and $u_{d\perp}$ the drifting speed of the electron beam. For Run3, $\gamma_{d}\approx 1.05$.

\begin{figure}[ht!]
  \plotone{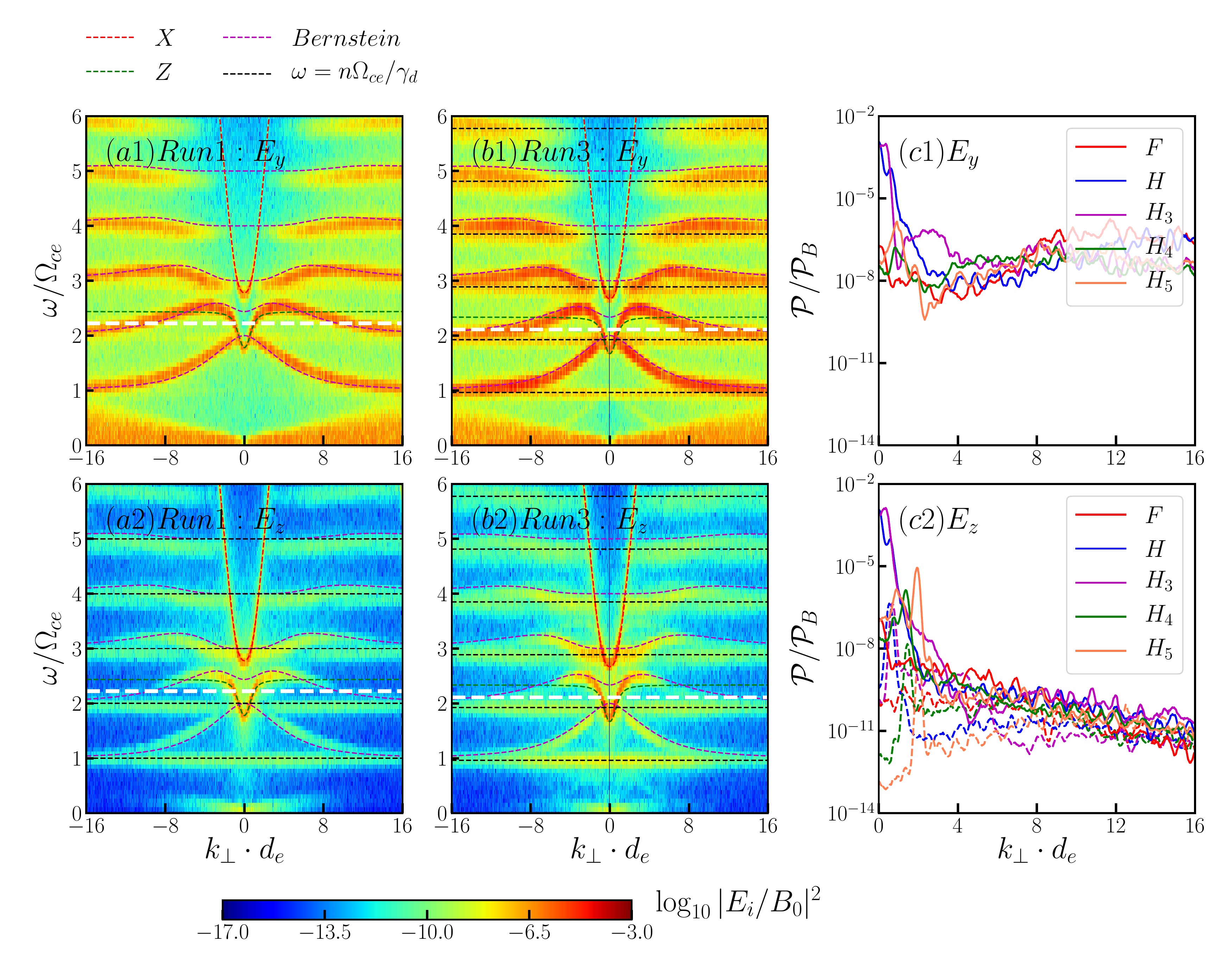}
  \caption{Power spectral density (PSD) for Run1 (a1--a2) and Run3 (b1--b2) in the time interval $t\cdot \omega_{pe}=455-634$. PSD estimated by $\log_{10}|E_i/B_0|^2\ (i=y,z)$ are displayed in the $k_{\perp}-\omega$ plane separately. Dispersion relation curves of X (red dashed curve), Z (green dashed curve), Bernstein modes (magenta dashed curves) and ECWs at frequency $\omega=n\Omega_{ce}/\gamma_{d}$ (black dashed curves) are overlaid. Note the equivalent Lorentz factor for Run1 (without beam) is $\gamma_{d}=1$. The white dashed line corresponds to local plasma frequency $\omega=\omega_{loc}$. Integrated power spectra $\mathcal{P}(k_{\perp})$ corresponding to the harmonics of ECWs $\omega=n\Omega_{ce}/\gamma_{d}$ are calculated in (c1) $E_y$ and (c2) $E_z$ for Run1 (dashed curves) and Run3 (solid curves) separately. The power spectrum $\mathcal{P}$ is normalized by initial magnetic field power density. \label{fig:EM_ECME}}
\end{figure}

\myreffig{fig:EM_ECME} (a1--a2,b1--b2) show the power spectral density, i.e., $\log_{10}|E_i/B_0|^2\ (i=y,z)$, in the $k_{\perp}-\omega$ plane for Run1 (without beam) and Run3 in the time interval $t\cdot \omega_{pe}=455-634$, respectively. EBWs are observed in PSDs for Run1 and Run3. Multiple harmonic gyroresonance-related ECWs are observed in PSD of $E_z$ for Run1, since the local plasma frequency is $\omega_{pe}\approx 2.2\Omega_{ce}$ (denoted by white dashed lines in panels of \myreffig{fig:EM_ECME}), only the third and higher ECWs may escape from the source region where they generated (see \myreffig{fig:EM_ECME}(a2)).
For Run3, a perpendicular crescent-shaped EVDF with a positive gradient in its 1D perpendicular EVDF, i.e., $\partial f(v_{\perp})/\partial v_{\perp}>0$, can offer sources of free energy to cause ECMIs and generate harmonics of electromagnetic ECWs (see \myreffig{fig:EM_ECME}(b1--b2)). Similarly, since the effective local plasma frequency is about $\omega_{loc}\approx 2.1\Omega_{ce}$ for Run3, only the third and higher harmonic of ECMI-generated ECWs may escape from the source region.

In order to quantitatively estimate the power spectral density of harmonics of ECWs, \myrefeq{eq:PSD_RS} is applied to calculate the power spectra for each harmonic of gyroresonance-related for Run1 and ECMI-generated ECWs for Run3 along their dispersion relation curves separately.
\myreffig{fig:EM_ECME} (c1) shows that the integrated power spectra of each harmonic of the ECMI-generated ECWs for Run3, while ECWs is absent in PSD of $E_y$ for Run1. For Run3, the magnitude of power spectra of ECWs is about $10^{-7}-10^{-2}$, the power spectra of the second to fourth harmonic ECWs are higher than that of the fundamental ECWs due to a wave coupling between EBWs and ECWs. In the long-wavelength regime, interaction between Z mode with the harmonic ECWs and X mode with the third harmonic ECWs enhance corresponding the power spectra of ECWs. The power spectra of the fifth ECWs is comparable to that of the fundamental ECWs. If its power spectra are strong enough, the third and higher harmonic ECWs may escape the source region where they are generated.

\myreffig{fig:EM_ECME} (c2) shows power spectra of ECWs derived from PSD of $E_z$. Power spectra of harmonics of the gyroresonance-related ECWs for Run1 are generally by order of $10^{-4}$ less than them of ECMI-generated ECWs for Run3. In particular, in the long-wavelength regime, power spectra of the second and higher harmonic are higher than that of the fundamental ECWs due to wave-wave interaction between Z and X modes with these ECWs for Run3. While in the short-wavelength regime, power spectra of the harmonic and higher harmonics of ECWs are slightly lower but generally comparable to them of the fundamental ECWs for Run3.

\begin{figure}[ht!]
  \plotone{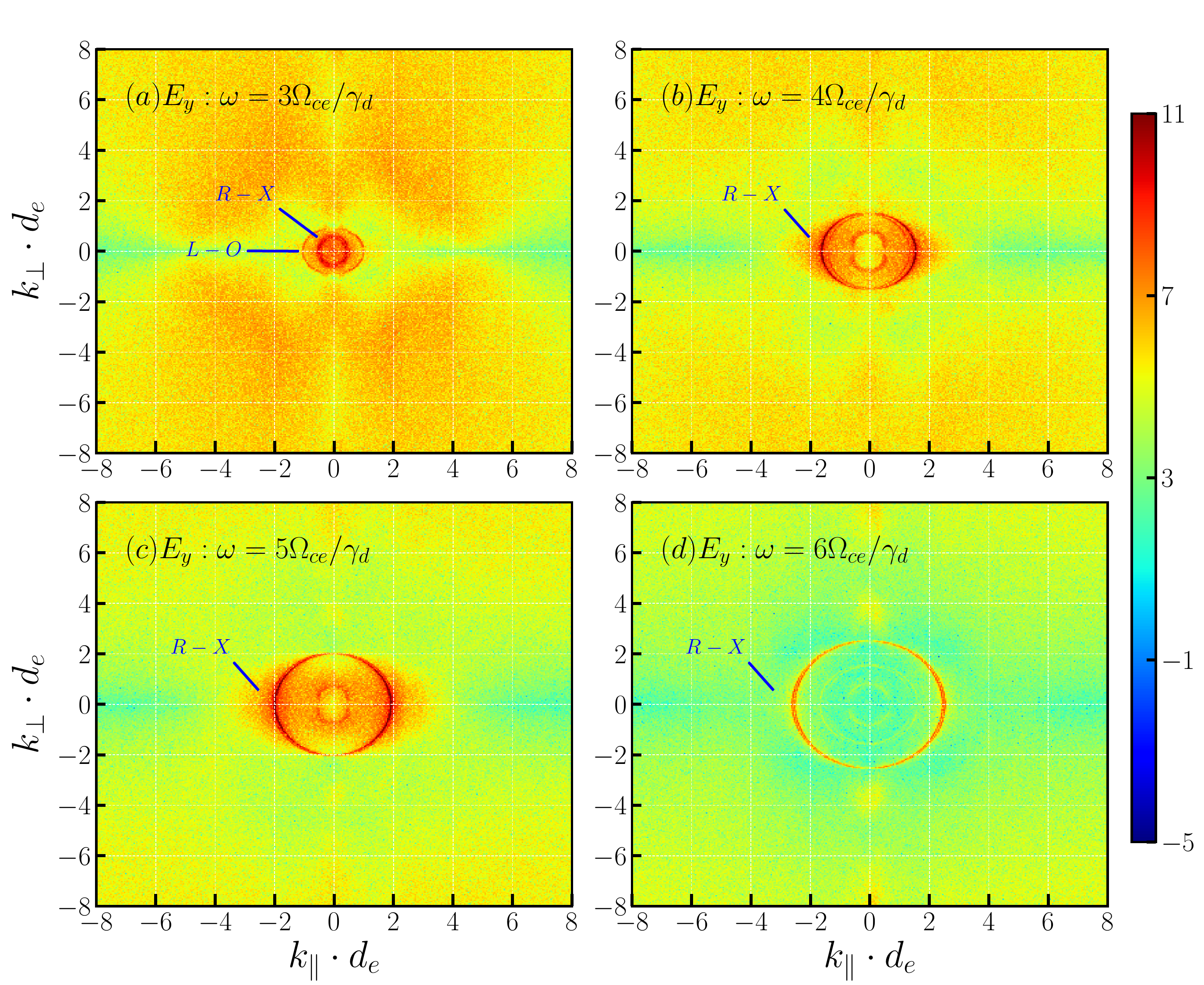}
  \caption{Power spectral density in the $k_{\parallel}-k_{\perp}$ plane of harmonics of ECWs at $\omega=3,\ 4,\ 5,\ 6 \Omega_{ce}/\gamma_{d}$, respectively. The PSD is estimated by $\log_{10}|E_y(\omega,k_{\parallel},k_{\perp})/B_0|^2$ at given $\omega$ for Run3 in the time interval $t\cdot \omega_{pe}=455-831$. When $\omega\ge 4\Omega_{ce}/\gamma_{d}$, the L-O mode approaches and merges with the R-X mode (denoted by the red circle at each panel).
  \label{fig:EM_ECME_3D}}
\end{figure}

\myreffig{fig:EM_ECME_3D} show PSDs of $E_y$ at $\omega=3,\ 4,\ 5,\ 6\Omega_{ce}/\gamma_{d}$, respectively, in $k_{\parallel}-k_{\perp}$ plane within time duration $t\cdot \omega_{pe}=455-831$ for Run3.
The radiation intensity of the third and fourth harmonic ECWs are nearly isotropic, while that of the fourth harmonic ECWs is slightly more intense along the perpendicular direction at $k_{\parallel}=0$ (see \myreffig{fig:EM_ECME_3D}(b)).
For the fifth harmonic ECWs, its radiation intensity is slightly stronger along the parallel direction about $k_{\perp}=0$ at $k_{\parallel}\cdot d_e=\pm 2$ (see \myreffig{fig:EM_ECME_3D}(c)), but radiation on the top-half and bottom-half plane are still significant.
The radiation intensity of the sixth harmonic ECW is weaker (see \myreffig{fig:EM_ECME_3D}(d)) than that of the lower harmonics of ECWs.

%%%%%%%%%%%%%%%%%%%%%%%%%%%%%%%%%%%%%%%%%%%%%%%%
% spectrogram
%%%%%%%%%%%%%%%%%%%%%%%%%%%%%%%%%%%%%%%%%%%%%%%%
\subsection{Transverse radio waves}\label{sec:transverse}

\myreffig{fig:EM_PE} (b1--b2) show the dispersion relation of transverse waves due to the electron/ion streaming of Run2.
Similarly,\myreffig{fig:EM_ECME} show the dispersion relation of transverse waves due to the ECMIs of Run3.
However, not all those waves can escape the plasma and be eventually remotely detected as radio waves. Only a small region of those diagrams represents radio emission.

\begin{figure}[ht!]
  \plotone{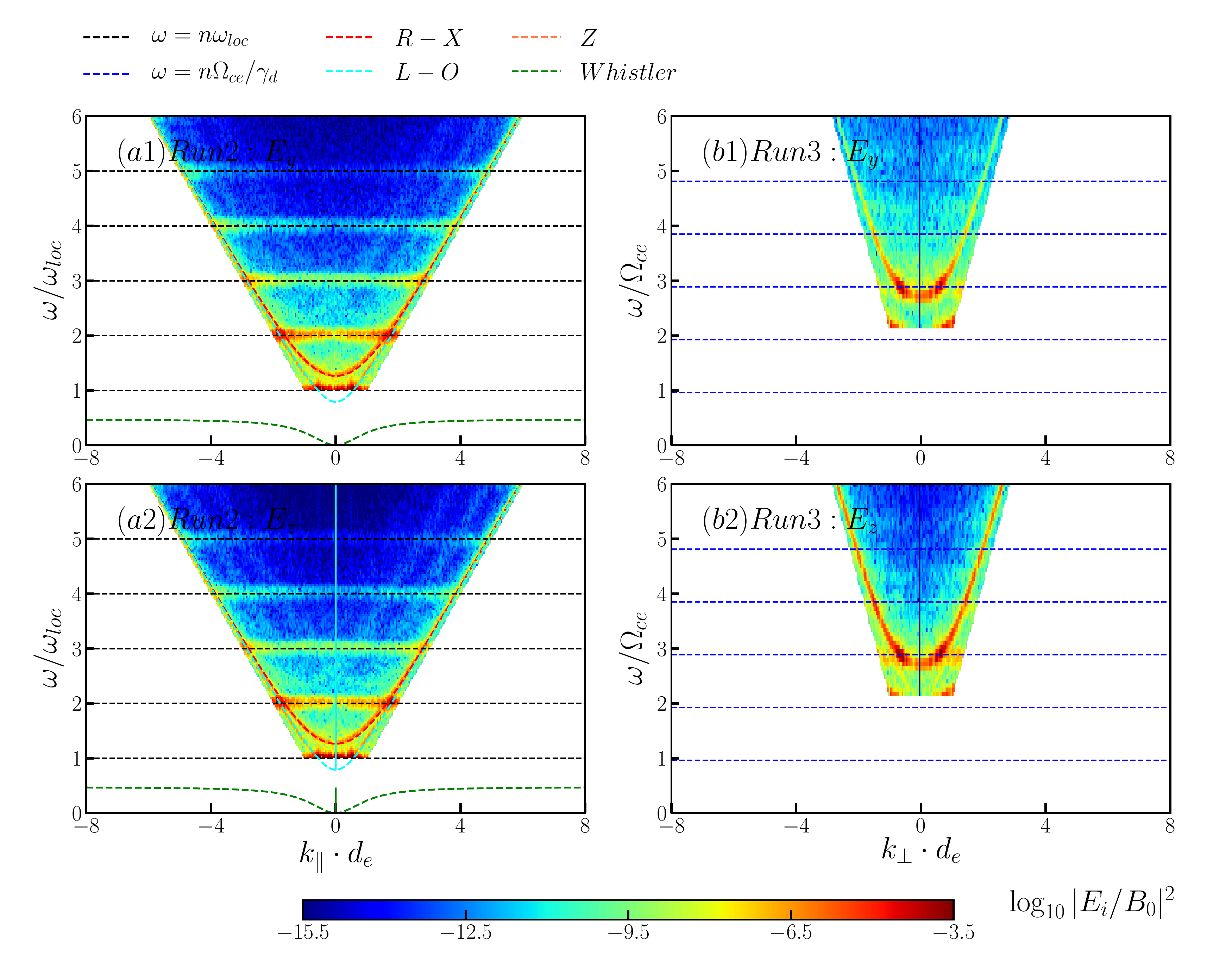}
  \caption{Power spectral density for the escaping electromagnetic transverse waves for Run2 (left) and Run3 (right). The PSD are derived form $E_y$ (top) and $E_z$ (bottom) component of the electric field respectively. Other parameters and labels are same to those shown in \myreffig{fig:EM_PE} and \myreffig{fig:EM_ECME}.}
  \label{fig:escaping_waves}
\end{figure}

Figure \ref{fig:escaping_waves} shows dispersion diagrams of transverse (electromagnetic) waves but constrained to two conditions: (1) waves' frequencies higher than the plasma frequency, i.e. $\omega>\omega_{pe}$, (2) superluminal waves, namely, whose phase speeds are larger than the speed of light, i.e., $\omega/k>c$. Those conditions are commonly known as ``escaping waves'' in the radio emission literature \citep{Melrose1986}. Figure \ref{fig:escaping_waves} shows the resulting dispersion relation of escaping electromagnetic waves of our simulations, which mostly follow the dispersion relation of the R-X mode with the presence of some harmonics of the plasma frequency for Run 2.

\begin{figure}[ht!]
  \plotone{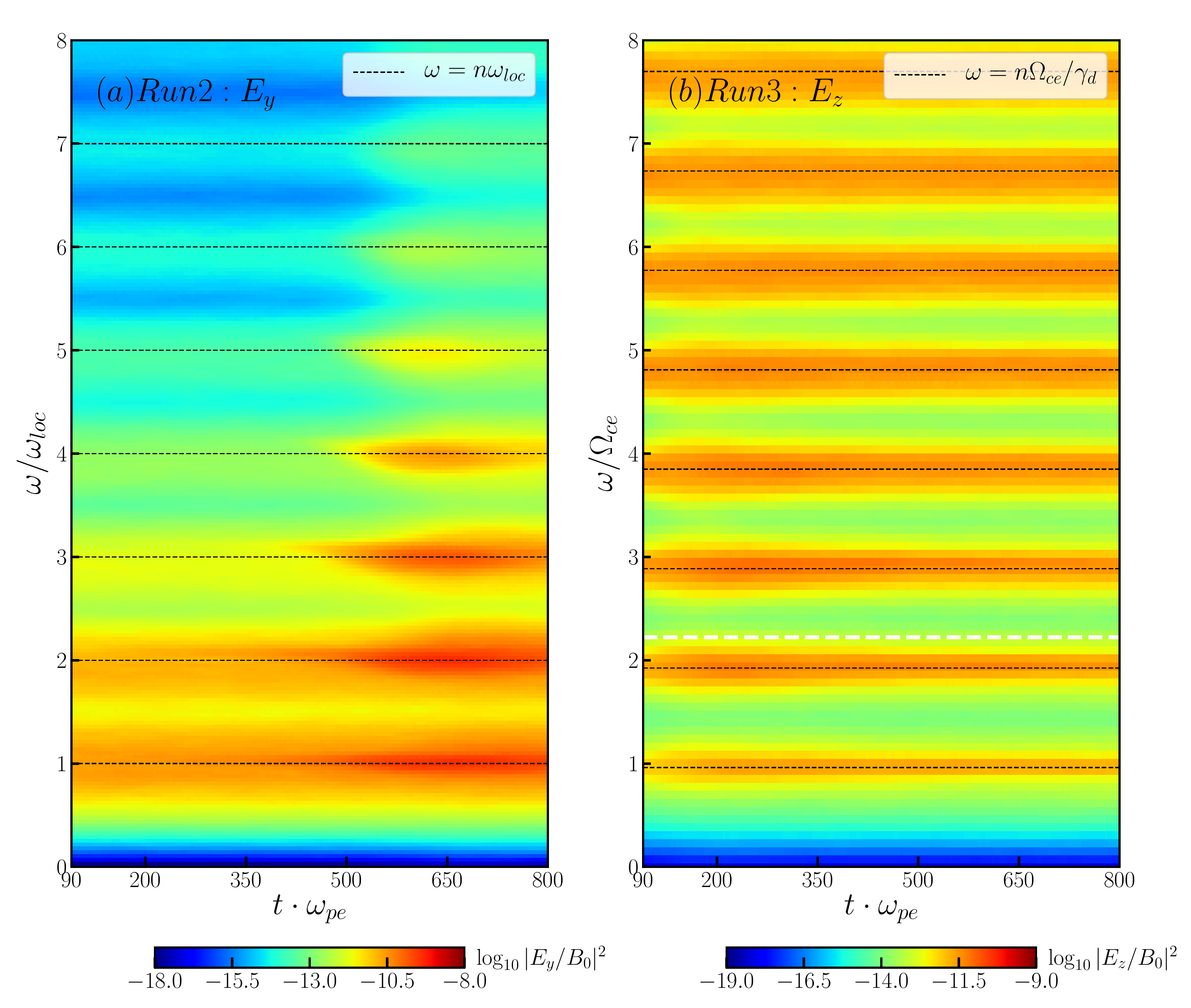}
  \caption{Spectrogram of harmonics of (a) transverse waves at frequency of $\omega=n\omega_{loc}$ for Run2 and (b) ECWs at frequency of $\omega=n\Omega_{ce}/\gamma_{d}$ for Run3. PSD estimated by $\log_{10}|E_{i}(t,\omega)/B_0|^2\ (i=y,z)$, respectively, is displayed in the $t-\omega$ plane, and the time window size is $\Delta T\cdot \omega_{pe}=179$. The black dashed lines indicate frequency locations by $\omega=n\omega_{loc}$ and $\omega=n\Omega_{ce}/\gamma_{d}$, separately. The white dashed line corresponds to $\omega=\omega_{loc}$. \label{fig:spectrogram_EM}}
\end{figure}

Another common diagnostic for electromagnetic emission in the radio wave range from the Sun is spectrograms (frequency--time diagrams). In particular for solar radio bursts. So in the following we present this diagnostics from our simulations.  The caveat of our simulated spectrogram is, of course,  that it is only due to the radiation emitted at the source region of radio bursts. Observed spectrograms of radio bursts follow the propagation of electron beams for very long distances compared to our simulations, actually comparable to the distance between the Sun and the Earth.

\myreffig{fig:spectrogram_EM} shows the spectrogram of transverse waves derived from $E_y(t,x)$ for Run2 and from $E_z(t,y)$ for Run3 in the $t-\omega$ domain. Note that for Run3 similar results are found in $E_y(t,x)$, here we just take results derived from $E_z(t,y)$ for example. The PSDs are calculated in the time window $t\cdot \omega_{pe}=90-800$ and with the window size $\Delta T\cdot \omega_{pe}=179$.

\myreffig{fig:spectrogram_EM}(a) shows significant features in the spectrogram of multiple harmonic plasma emission due to electron/ion streaming instabilities. They are summarized as follows:
\begin{itemize}
    \item Harmonics of transverse waves, e.g., up to $6th$ harmonic, are generated, even with (PSD of) the $7th$ and $8th$ harmonic in a relatively weak level.
    \item All harmonic components of the transverse waves can be generated in the same sources region at short electron time scale $t\cdot \omega_{pe}\le 500$.
    \item The fundamental and harmonic transverse waves are generated earlier than the third and higher harmonics of plasma waves, e.g., the fundamental and harmonic waves are produced at $t\cdot \omega_{pe}=90$, while the third to $6th$ harmonic waves occur after $t\cdot \omega_{pe}=500$.
\end{itemize}

Note that  \myreffig{fig:spectrogram_EM}, in combination with our previous discussions, also provides evidence for the role of ion-acoustic waves in the generation of
electromagnetic transverse emission.
Indeed, this figure shows that higher harmonic emission as well as stronger fundamental and first harmonic emission only occur after $t\cdot \omega_{pe}\approx 400-500$.
This agrees with the exponential growth of transverse electromagnetic wave energy (seen in \myreffig{fig:growth_rate_3runs}(b)), as well as with the growth of longitudinal
electrostatic wave energy (see in \myreffig{fig:growth_rate_3runs}(a)) at this same time period.
Note that longitudinal electrostatic wave energy includes mainly (forward
and backward) propagating Langmuir waves as well as ion-acoustic waves.
The growth of the latter in form of ion density fluctuations can be clearly seen also at the same time (see in \myreffig{fig:IA_waves}(a)).
All those findings are also in agreement with our linear theory calculations (based on the instantaneous distribution functions) that predict  unstable exponentially growing waves in the ion-acoustic branch only after this time period (see \myreffig{fig:VDFs_growth_rate}).
In summary, those results indicate that there is a correlation between ion-acoustic waves
and generation of electromagnetic wave by the plasma emission mechanism. Note that this is not necessarily a causal relation. It just provides evidence in favor of this process.

{\bfseries
An additional evidence for the relation between ion-acoustic waves $S$ and generation of electromagnetic emission is the wave-wave matching (beating) conditions which are a manifestation of the conservation of momentum (for the wavenumber) and energy (for the frequency).
Here we focused on one possible pathway of wave-wave interaction involving ion-acoustic waves $S$ and electrostatic Langmuir waves $L$.
First we provide simulation evidence for the following standard interaction:
\begin{align}
  L-S\to F
  \label{eq:wave-wave}
\end{align}
here $F$ represents the fundamental electromagnetic emission at frequency $\omega=\omega_{pe}$. This can be interpreted as a decay process of $L$ waves in the form $L\to S+F$  (discussed in the introduction), but somehow different because the driver are ion-acoustic waves. So it is not exactly the same process reported before in the literature \citep[see, e.g.,][]{Thurgood2015}. 

The decay process \myrefeq{eq:wave-wave} implies the following beat conditions for frequency and wavenumber:
\begin{align}
  k_{L} - k_{S} &= k_F\label{eq:wave-wave-k}\\
  \omega_L-\omega_S&=\omega_F\label{eq:wave-wave-w}
\end{align}

In general \myrefeq{eq:wave-wave-k} involves all components of the wavenumber vector (see \myreffig{fig:EM_PE_3D} for the angular pattern of wave emission) but for the sake of simplicity we will limit ourselves to just the parallel direction (i.e. $x$ direction). 

For ion-acoustic waves $S$, the wavenumber $k_{S}$ can be obtained from \myreffig{fig:IA_waves} (b). The region with significant (much higher than the surrounding thermal noise) spectral power along the dispersion relation curve of ion-acoustic waves is roughly confined to $k_{\parallel}\cdot d_e\in[0.4, 4]$ after $t\cdot \omega_{pe}\approx 400$.
An analysis of the integrated 1D spectra $k_{\parallel}$ as a function of time (not shown here) shows that the spectral power is of course not constant in time: it broadens from lower to larger wavenumbers as the time evolves. There is a broad spectral peak that also shifts in time, but it is roughly located between $k_{\parallel}\cdot d_e\in[0.4, 1.5]$. This roughly agrees with the most unstable wavenumbers at the beginning of the exponentially growing phase of the instability (see \myreffig{fig:VDFs_growth_rate}(b2)). Therefore, the wavenumber of ion-acoustic waves is $k_{S}\cdot d_e\in[0.4,1.5]$, and the corresponding frequency can be similarly obtained from \myreffig{fig:IA_waves} (b) or be quantitatively estimated by \myrefeq{eq:disp_rela_S}.

For the Langmuir waves $L$, the wavenumber $k_{L}$ can be obtained from \myreffig{fig:EM_PE}(b1) and \myreffig{fig:EM_PE}(c1). The red line in \myreffig{fig:EM_PE}(c1) shows that most of the power is concentrated between $k_{\parallel}\cdot d_e=0$ and $k_{\parallel}\cdot d_e=0.75$ (both of which represent a local maxima), although strong power extends toward significant higher wavenumbers (up to approximately $k_{\parallel}\cdot d_e=3$). Note that the spectral power is directionally asymmetric: Langmuir waves with negative $k$ are a bit stronger and extend to larger negative $k$ than Langmuir waves with positive $k$. The corresponding frequency $\omega_L$ can be obtained from \myreffig{fig:EM_PE}(b1), or be quantitatively estimated by the standard Bohm-Gross dispersion relation mentioned above.

For the fundamental emission, $k_F$ can be obtained from \myreffig{fig:EM_PE}(b2) and \myreffig{fig:EM_PE}(c2). There is a clear spectral power peak at the intersection of $\omega=\omega_{pe}$ with the electromagnetic mode (green dashed curve in \myreffig{fig:EM_PE}(b2), located at $k_{\parallel}\cdot d_e=0.48$. However, there is still significant power in the wavenumber range  $k_{\parallel}\cdot d_e\in[0,1]$ that comes from the projection to $k_{\perp}=0$ due to spectral power emitted at oblique angles (see \myreffig{fig:EM_PE_3D}(b)). The spectral power outside the electromagnetic mode but around $\omega=\omega_{pe}$, which is not described by linear theory, is a manifestation of its non-linear nature.

Note the wavenumber and frequency resolution for the spectral power of ion density fluctuations (shown in \myreffig{fig:IA_waves} (b)) is $\Delta k=0.044d_e^{-1}$ and $\Delta \omega=0.0044\omega_{loc}$ (with 4096 samplings in time) separately, while for the power spectral density of electric fields (shown in \myreffig{fig:EM_PE}) is $\Delta k=0.044d_e^{-1}$ and $\Delta \omega=0.035\omega_{loc}$ (with 512 samplings in time) separately.

This way, one possible interaction leading to the fundamental emission as per \myrefeq{eq:wave-wave-k} and \myrefeq{eq:wave-wave-w} is:
\begin{align}
  (k_L d_e=0.96\pm0.04) - (k_S d_e=1.45\pm0.04) &\to (k_F d_e=-0.48\pm0.04)\label{eq:neg_F_k}\\
  (\omega_L/\omega_{loc}=1.02\pm 0.035) - (\omega_S/\omega_{loc}=0.022\pm 0.004) &\to (\omega_F/\omega_{loc}=0.98\pm 0.035)\label{eq:neg_w}
\end{align}
Note that this leads to a spectral peak in the electromagnetic wave dispersion relation curve with anti-parallel propagation (negative $k$).
The ambiguity in this equation is mainly contained in $k_S$ (i.e., one of many possible values), since the spectral power in ion-acoustic waves is broadly distributed (while the Langmuir and fundamental emission peaks are well defined).
Frequencies for each wave candidate are $\omega_L=(1.02\pm 0.035)\omega_{loc}$ for Langmuir waves, $\omega_S=(0.022\pm 0.004)\omega_{loc}$ for ion-acoustic waves, and $\omega_F= (0.98\pm 0.035) \omega_{loc}$ for fundamental emission respectively (see \myrefeq{eq:neg_w}). Note the errors due to the spectral resolution are relatively large, which could explain the slight difference between the calculated $\omega_F$ and  $\omega_{loc}$.

The complementary spectral peak (in the dispersion relation curve of fundamental emission  $\omega=\omega_{loc}$) with parallel propagation (i.e. positive $k_{\parallel}$) can be obtained from counter-propagating Langmuir and ion acoustic-waves (often referred in the literature as a coalescence process $L'+S\to F$), which have spectral peak values with similar wavenumber magnitudes but negative sign as per \myreffig{fig:IA_waves}(b) and \myreffig{fig:EM_PE}(b1), namely
\begin{align}
  (k_L d_e=-0.96\pm 0.04) - (k_S d_e=-1.45 \pm 0.04) &\to (k_F d_e=0.48 \pm 0.04)
  \label{eq:pos_F}
\end{align}
These somehow symmetric spectral peaks of fundamental emission at $k_F d_e=\pm 0.48$, namely, positive and negative $k_{\parallel}$ estimated by Eqs. \eqref{eq:pos_F} and \eqref{eq:neg_F_k}, resemble the results obtained by \citet{Ganse2012}. They are, however, different than the strongly peaked with only positive $k$ fundamental emission observed in the simulations by \citet{Thurgood2015}. One reason for those discrepancies may lie on the very different parameter set and distribution functions used for these simulations.

The spectral power at the fundamental emission near $k_Fd_e=0$ (see \myreffig{fig:EM_PE}(2) and \myreffig{fig:EM_PE}(c2)), and in general in the region in between the peaks at the electromagnetic mode could also be generated as a consequence of a wave-wave interaction between Langmuir and ion-acoustic waves. For example,
\begin{align}
  (k_L d_e=0.96\pm 0.04)-(k_S d_e=0.96\pm 0.04) &\to (k_F d_e=0\pm 0.04)\\
  (\omega_L/\omega_{loc}=1.02\pm 0.035)-(\omega_S/\omega_{loc}=0.013\pm 0.004)&\to(\omega_F/\omega_{loc}=1.05\pm 0.035)\label{eq:LS_F_w}
\end{align}
Again, the mismatch between $\omega_F=(1.05\pm 0.035)\omega_{loc}$ and $\omega_{loc}$ could be explained by the finite spectral resolution.
In general, fundamental emission in the wavenumber range $k_F d_e\in[-0.48, 0.48]$ could originate primarily as a result of similar interactions with ion-acoustic waves in corresponding range $k_Sd_e\in[0.4, 1.5]$ within which $S$ waves are broadband with higher power. Langmuir waves could have been also chosen at wavelengths different from $k_L d_e=\pm0.48$, since they are broadband all the way up to $k_L d_e=0$, leading to similar possible interactions.
Frequencies for each wave candidate are $\omega_L=(1.02\pm 0.035)\omega_{loc}$ for Langmuir waves, $\omega_S=(0.013\pm 0.004)\omega_{loc}$ for ion-acoustic waves, and $\omega_F= (1.05\pm 0.035)\omega_{loc}$ for fundamental emission respectively (see \myrefeq{eq:LS_F_w}).

Counter-propagating Langmuir waves can be generated by means of the process $L+S\to L'$. For example,
\begin{align}
  (k_L d_e=0\pm 0.04) + (k_S d_e=-0.96\pm 0.04) &\to (k_{L'} d_e=-0.96\pm 0.04)\\
  (\omega_L/\omega_{loc}=1.02\pm 0.035)+(\omega_S/\omega_{loc}=0.013\pm 0.004)&\to(\omega_{L'}/\omega_{loc}=1.033\pm 0.035)\label{eq:LS_Lback}
\end{align}
It is also clear that higher (and broadband) spectral power in the $S$ waves  implies generation of counter-propagating Langmuir waves $L'$. \myrefeq{eq:LS_Lback} shows frequencies for each wave candidate are $\omega_L=(1.02\pm 0.035)\omega_{loc}$ for Langmuir waves, $\omega_S=(0.013\pm 0.004)\omega_{loc}$ for ion-acoustic waves, and $\omega_{L'}=(1.033\pm 0.035)\omega_{loc}$ for back-scattered Langmuir waves respectively.

Another important remark is that this wave-wave interaction may in principle always occur in a plasma where those $L$ and $S$ waves are excited, even without a beam, like in our Run1. Those modes are just linear plasma waves, but with very low amplitude and, in the case of ion-acoustic waves, highly damped. Therefore, any kind of fundamental emission is negligible in this case. This absence of emission occurs also in the initial stage of our simulation ($t\omega_{pe}<400 $), since ion-acoustic waves are damped and only grow later (see \myreffig{fig:IA_waves}).
The wave-wave interactions efficiently occur after this period ($t\omega_{pe}>400 $), as evidenced by the increased power at the fundamental mode in the spectrogram of \myreffig{fig:spectrogram_EM}(a).

The enhanced fundamental electromagnetic emission mode plus forward- and backward-propagating Langmuir waves after $t \omega_{pe} > 400$ are the seed for the harmonic (and higher harmonic) electromagnetic emission that appears after this time (see \myreffig{fig:spectrogram_EM}(a)), as a consequence of the interaction $L+L'\to H$.  Since Langmuir waves are weak before this time, $H$ emission is barely observed. Note that finding a combination of waves that corresponds to the spectral peaks of $H$ near the electromagnetic mode is not straightforward, although it could be attributed to the broadband nature of Langmuir waves.
}

Similar features for multiple harmonic ECWs due to ECMIs are displayed in \myreffig{fig:spectrogram_EM}(b). However, all harmonics of ECWs are simultaneously formed in the source region, i.e., $t\cdot \omega_{pe}\le 90$.
Higher harmonics of ECWs are generated earlier than their counterparts of plasma emission caused by two-streaming EVDFs.
For example, the third to higher harmonic of ECWs can be already generated within $t\cdot \omega_{pe}=0-500$, while their plasma emission counterparts have not yet formed. Different from the ECWs caused by ECMIs, \myreffig{fig:spectrogram_EM}(a) shows that the $4th$ to $6th$ harmonic of transverse plasma waves (caused by electron/ion streaming instabilities) reach their maximum in $t\cdot \omega_{pe}=520-700$ and are damped significantly after $t\cdot \omega_{pe}=700$. Because multiple harmonic plasma emission is attributed to a series of multi-stage wave--wave interaction process as discussed before, the energy density of higher harmonic plasma emission comes from the lower harmonics. As a result, the energy density in higher harmonic plasma emission is lower than that in the lower harmonics.

%%%%%%%%%%%%%%%%%%%%%%%%%%%%%%%%%%%%%%%%%%%%%%%%
%Conclusions
%%%%%%%%%%%%%%%%%%%%%%%%%%%%%%%%%%%%%%%%%%%%%%%% 
\section{Summary and conclusions}\label{sec:conclusions}

By virtue of fully-kinetic PIC code simulations, we investigated the generation of unstably growing plasma oscillations and the excitation of radio waves due to non-thermal electron velocity space distribution functions (EVDFs) generated by 3D kinetic magnetic reconnection. The reconnection generated non-thermal EVDFs have previously been found in numerical experiments of PIC simulations and by {\it in situ} observations.

A fully self-consistent kinetic simulation of radio emissions by magnetic reconnection cannot be achieved due to the still insufficiently low temporal and spatial resolution of 3D kinetic magnetic reconnection simulations.
Instead, we first carried out 3D PIC code simulations of kinetic reconnection with a relatively sparse output in time, which is still computationally expensive, to generate the characteristic non-thermal EVDFs \citep{Yao2022}.
Using them, we have carried out 2.5D PIC code simulations of beam-plasma interaction, allowing a sufficiently high frequency and wavenumber resolution in the Fourier domain to derive the properties of resulting radio waves.
As already explained in \citep{Yao2022}, the magnetic reconnection simulation does not show significant signatures of streaming instabilities, mainly due to the fact that the possible streaming instabilities are relatively weak in comparison with other macroscopic instabilities and plasma flows in magnetic reconnection. Nevertheless, numerical simulations of a beam-plasma system  with higher spectral resolution allows analysis of the possible radio emission caused by streaming instabilities.

In this way, we bridged the gap between the non-thermal EVDFs generated in magnetic reconnection and the formation of radio waves caused by the characteristic non-thermal EVDFs offering source of free energy. Therefore, remote observation of the radio waves, e.g., during solar flares, can be used as a tool for remote diagnostics of magnetic reconnection in astrophysical plasmas.

We first investigated the consequences of multiple harmonic plasma waves caused by two-streaming EVDFs formed along the separatrix and in the diffusion region of 3D kinetic magnetic reconnection, then we investigated the consequences of multiple harmonic electron cyclotron waves caused by perpendicular crescent-shaped EVDFs generated in the diffusion region of 3D kinetic magnetic reconnection \citep{Yao2022}.

The two-streaming EVDFs (Run2) can offer sources of free energy to cause electron/ion streaming instabilities, which can generate ion acoustic waves.
The relative electron/ion streaming can also generate Langmuir waves.
Those EVDFs are found in the separatrices and diffusion region of kinetic reconnection.
Multiple harmonic transverse plasma waves at the frequency of multiples of the local electron plasma, i.e., $\omega=n\omega_{loc}$, due to non-linear wave--wave interaction processes are generated. The excitation of multiple harmonic plasma waves is sensitively dependent on the field-aligned drift speed of the electron beam and on the beam-to-background electron number density ratio $n_{bm}/n_{bg}$. For small field-aligned electron drift speed, the beam-to-background electron number density ratio $n_{bm}/n_{bg}$ must be large enough in order to excite harmonic(s) of plasma emission.
Thus, observations of harmonic emission would allow to constraint the values of both field-aligned drift speed and beam-to-background electron number density ratio of the electron beam.

We then investigated perpendicular crescent-shaped EVDFs (Run3) generated near and in the diffusion of reconnection. Such kinds of EVDFs are able to offer sources of free energy to cause electron cyclotron maser instabilities, which can generate multiple harmonic electromagnetic electron cyclotron waves  due to a wave--particle interaction. After being amplified by wave--particle interactions, the latter can be isotropically emitted in all directions.
Remote observations of electromagnetic electron cyclotron waves can be used for the remote diagnostics of the magnetic field as well as of the ratio of the electron cyclotron frequency to the electron plasma frequency in reconnection regions.

In both cases, the EVDFs of fast electron beams contain a two-streaming EVDF in the field-aligned direction. This results in the formation of multiple harmonic plasma emission.

Radiation intensities of harmonics of plasma emission are anisotropically distributed and markedly angular-dependent \citep[][]{Rhee2009,Zlotnik1978,Reiner1992,Takakura1974}. The higher the harmonic number of the plasma emissions, the closer the distribution of radiation intensity approaches the field-aligned direction. Our findings have a significant practical application on remote diagnostics of reconnection by observations of harmonics of transverse emission, namely, due to the fact that the radiation intensity of harmonic of plasma emission heavily depends on the observation angle and the relative intensity variation in different locations.

Based on the angular pattern of radiation intensity of multiple harmonic plasma emission, our results can also serve as a theoretical testing ground for verifying different processes of the formation of each harmonic plasma emissions (see Appendix \ref{app:Radiation_pattern}). 
Harmonics of electrostatic Langmuir waves are excited through a coalescence process $L+L_{n-1}\to L_n$ \citep[][]{Yi2007,Rhee2009,Yao2021}. 
Via a process mediated by ion-acoustic waves $S$, harmonics of transverse emission at a frequency of multiples of the plasma frequency can be generated by non-linear wave--wave interaction processes as follows:

\begin{itemize}
    \item Fundamental transverse emissions $F$ can be generated by a decay process of electrostatic Langmuir waves $L$ and ion-acoustic waves $S$, i.e., $L\to S+F$ \citep[][]{Robinson1994a,Reid2014,Melrose2008}. 
    \item Harmonic transverse emissions $H$ can be generated by a coalescence process of electrostatic Langmuir waves $L$ and backward scatted Langmuir waves $L'$ via $L+L'\to H$ \citep[][]{Willes1996a,Yoon2006,Rhee2009}. 
    \item Third and higher harmonics of transverse emission $H_n\ (n=3,4,5...)$ can be generated by a coalescence process of electrostatic Langmuir waves $L$ and transverse emission $H_{n-1}$, i.e., $L+H_{n-1}\to H_n$ \citep[][]{Zlotnik1978,Cairns1988,Rhee2009}.
\end{itemize}

Based on the above discussion, our findings can be used for the remote radio diagnostics of magnetic reconnection by which non-thermal EVDFs are generated and subsequently cause radio waves, in particular in the course of solar flares. 
Note that radio waves and emission generated directly in the site where magnetic reconnection takes place were observed. For example, observations of fundamental and harmonic plasma emission near the null point of reconnection in solar flares \citep[][]{Chen2018}.

We did not consider propagation effects of the electron beam in the ambient plasma, like in the solar corona or in the solar wind, on the properties of radio waves. They should be taken into account for the sake of a correct interpretation of the observed radio signals. We just investigated the initial stage of the emission process.

Another small caveat is that we are only considering the EVDFs from magnetic reconnection as initial conditions for the beam-plasma simulations, in addition to similar parameters.
We do not use other plasma conditions resulting from the magnetic reconnection evolution, e.g., density gradients, partially for simplicity but also because those effects have already been investigated \citep{Yao2021}.
In this same sense, in our beam-plasma simulations we do not initially impose any of the additional waves caused by the reconnection process.
As it is well-known, not only gradients in density or bulk flow generated by reconnection can cause waves, but also Hall currents which lead to streaming instabilities, temperature-anisotropy driven instabilities, etc. Those plasma micro-instabilities can finally generate Langmuir waves, whistler waves, kinetic Alfv\'en waves, etc \citep{Fujimoto2011,Treumann2013,Khotyaintsev2019}.
Some of those waves are caused by the EVDFs, but they can also perturb and modify the EVDFs themselves, in particular high-frequency waves with phase speed near its tail via resonant interactions.
Both the waves and the distribution function must obey the Vlasov-Maxwell system of equations, which is coupled, and therefore there is always an interplay between electromagnetic fields and distribution functions. One of the most important effects of waves on EVDFs is their apparent perpendicular heating and consequent anisotropic heating, which is different from ``real'' heating in that it is reversible \citep{Verscharen2011}.
Since waves and EVDFs are intrinsically linked, it is hard to disentangle the individual effects of the waves caused by the EVDFs from the waves that perturb and contribute to the formation of the EVDFs.
So we decide in the present work to analyze the EVDFs and waves caused by them, not the waves that could actually lead to the formation of EVDFs in reconnection.

%%%%%%%%%%%%%%%%%%%%%%%%%%%%%%%%%%%%%%%%%%%%%%%%
% Acknowledgements
%%%%%%%%%%%%%%%%%%%%%%%%%%%%%%%%%%%%%%%%%%%%%%%%
\section{Acknowledgements}

We gratefully acknowledge the developers of the ACRONYM code, the \textit{Verein zur F\"orderung kinetischer Plasmasimulationen e.V.} and the financial support by the German Science Foundation (DFG), projects MU-4255/1-1, BU 777/15-1, and BU~777-17-1.
We also acknowledge the Sino-German collaboration made possible thanks to the projects with National Natural Science Foundation of China (NSFC) grant 11761131007 and 12003073 and with the Internaltional Scholarship Program of the MOST of China G2021166002L.
We also gratefully acknowledge the possibility of using the computing resources of the Max Planck Computing and Data Facility (MPCDF, formerly known as RZG) at Garching and of the Max-Planck-Institute for Solar System Research at G\"ottingen as well as of the Technical University Berlin, Germany. 
We thank the referee for their useful remarks that have helped us to improve the original version of this paper.

%%%%%%%%%%%%%%%%%%%%%%%%%%%%%%%%%%%%%%%%%%%%%%%%
% Declaration of interests
%%%%%%%%%%%%%%%%%%%%%%%%%%%%%%%%%%%%%%%%%%%%%%%%
\section{Declaration of interests}
The authors report no conflicts of interest.

%%%%%%%%%%%%%%%%%%%%%%%%%%%%%%%%%%%%%%%%%%%%%%%%
% Appendix
%%%%%%%%%%%%%%%%%%%%%%%%%%%%%%%%%%%%%%%%%%%%%%%%
\appendix
\section{Radiation pattern of multiple harmonic plasma emission}\label{app:Radiation_pattern}

\cite{Rhee2009} proposed that for each harmonic of transverse emission the associated wave-wave interaction processes can be verified by the angular pattern of its radiation intensity. Based on previous studies of radiation intensity pattern \cite[e.g., see][]{Rhee2009,Yoon2006}, we extended to talk about angular pattern of radiation intensity of multiple harmonics of plasma emission.

For the fundamental transverse emission $F$, it is generated by a decay process of electrostatic Langmuir waves $L$ and ion-acoustic waves $S$, i.e., $L\to S+F$.
It is intuitively understood that the angular pattern of fundamental emission should be $90^{\circ}$ different from that of the electrostatic Langmuir waves $L$.

An theoretical analysis of the amplification of $F$ emission due to $L\to S+F$ can be estimated by the transverse wave kinetic equation \citep[][]{Yoon2005a,Yoon2005,Yoon2006} as follows:
\begin{eqnarray}
  \frac{\partial I_F(\vec{k})}{\partial t}\propto\frac{\pi e^2}{m_e^2T_e^2}\int d\vec{k'}\frac{\left(\vec{k}\times\vec{k'}\right)^2}{k^2k'^2|\vec{k}+\vec{k'}|^2} \cdot I_L(\vec{k'})I_S(\vec{k}+\vec{k'})\delta\left(\omega_{F}(\vec{k})-\omega_{L}(\vec{k'})-\omega_{S}(\vec{k}+\vec{k'})\right)
\end{eqnarray}
here $I_L(\vec{k})$ and $I_F(\vec{k})$ are the spectral intensities of electrostatic Langmuir waves $L$ and fundamental emission $F$ separately.
Assume isotropic electrostatic Langmuir wave intensities, i.e., $I_L(\vec{k})\equiv const$, then the radiation angular pattern for the fundamental emission $F$ is determined by the transition probability term, i.e., $\left(\vec{k}\times\vec{k'}\right)^2\sim \sin^2\theta$.
Here $\theta$ is the wave phase angle in the $k_{\parallel}-k_{\perp}$ plane. As a result, the radiation intensity of $F$ emission mainly occurs in angle range about $90^{\circ}$ in the top-half plane and $270^{\circ}$ in the bottom half-plane, as that shown in \myreffig{fig:EM_PE_3D}(b).
Note that the radiation pattern of the electrostatic langmuir waves is, however, strongly angle dependent, i.e., at angles $0^{\circ}$ and $180^{\circ}$ (see \myreffig{fig:EM_PE_3D}(a)).
Nevertheless, the isotropic assumption is still valid because radiation of the electrostatic Langmuir waves is very strong elsewhere \citep[][]{Rhee2009}.

For the harmonic emission $H$, as we mentioned above, it is generated by a coalescence process of beam-generated Langmuir waves $L$ and backward scattered Langmuir waves $L^{\prime}$, i.e., $L+L^{\prime}\to H$ \citep[][]{Willes1996a,Yoon2006}.
The amplification of $H$ radiation associated with this coalescence process can be evaluated by the transverse wave kinetic equation \citep[][]{Yoon2005a,Yoon2005,Yoon2006}
\begin{eqnarray}
    \frac{\partial I_H(\vec{k})}{\partial t}\propto\frac{\pi e^2}{m_e^2\omega_{pe}^2}\int d\vec{k'}\frac{\left(\vec{k}\times\vec{k'}\right)^2}{k^2k'^2}\frac{|k^2-2\vec{k}\cdot\vec{k'}|^2}{|\vec{k}-\vec{k'}|^2} \cdot I_L(\vec{k'})I_L(\vec{k}-\vec{k'})\delta\left(\omega_{H}(\vec{k})-\omega_{L}(\vec{k'})-\omega_{L}(\vec{k}-\vec{k'})\right)
    \label{eq:amplification_EM_H}
\end{eqnarray}
here $I_H(\vec{k})$ is the spectral intensities of harmonic emission $H$. For isotropic electrostatic Langmuir wave intensity, the transition probability term, i.e., $\left(\vec{k}\times\vec{k'}\right)^2|k^2-2\vec{k}\cdot\vec{k'}|^2\sim\sin^2\theta\cos^2\theta$, determines the angular pattern for radiation intensity of the harmonic emission $H$. The harmonic emission $H$ shows the quadrupole pattern of radiation intensity, which mainly occurs in each quadrant in the $k_{\parallel}-k_{\perp}$ plane, i.e., in angle range about $\theta=45^{\circ}, 135^{\circ}, 225^{\circ}$ and $315^{\circ}$, see \myreffig{fig:EM_PE_3D} (c).

For the third and higher harmonic of transverse emission, \cite{Rhee2009} pointed out that they are generated by a coalescence process of electrostatic Langmuir waves $L$ and another transverse wave mode $T$. An candidate model satisfies this requirement is $L+H_{n-1}\to H_n$, which means the nth harmonic emission $H_n$ is generated by a coalescence process of electrostatic Langmuir waves $L$ and transverse emission $H_{n-1}$ \citep[][]{Zlotnik1978,Cairns1988}.

We represent the wave vector of the nth transverse emission $H_n$ involved in the coalescence process $L+H_{n-1}\to H_n$ as $\vec{k}_n \to (k_n,\theta_n)$ and electrostatic Langmuir waves as $\vec{k}_L \to (k_L,\theta_L)$. Here the wave propagation angle $\theta_n$ is defined as the angle between the direction of local magnetic field and that of wavevector $\vec{k}_n$. 

They satisfy the beat condition $\vec{k}_L+\vec{k}_{n-1}=\vec{k}_{n}$. Thus the wave vector for $L$ is $\vec{k}_L=\vec{k}_n-\vec{k}_{n-1}$ and the magnitude of vector $\vec{k}_n-\vec{k}_{n-1}$ lies within the range
\begin{equation}
    k_n-k_{n-1}\le|\vec{k}_n-\vec{k}_{n-1}|\le k_n+k_{n-1}
    \label{eq:kL_range}
\end{equation}
where $k_n=\sqrt{n^2-1}\cdot\omega_{pe}/c$ is the magnitude of vector $\vec{k}_n$, it can be solved from dispersion relation of transverse wave mode for transverse emission $H_n$ \citep[][]{Melrose1986}.

The cosine of the angle of vector $\vec{k}_L=\vec{k}_n-\vec{k}_{n-1}$ therefore reads
\begin{equation}
    \cos\theta_L=\frac{k_n\cos\theta_n-k_{n-1}\cos\theta_{n-1}}{|\vec{k}_n-\vec{k}_{n-1}|}
\end{equation}

Taking \myrefeq{eq:kL_range} into account, yielding the following inequation
\begin{equation}
    \min\left(f_1(\theta_n,\theta_{n-1}),f_2(\theta_n,\theta_{n-1})\right)\le cos\theta_L \le \max\left(f_1(\theta_n,\theta_{n-1}),f_2(\theta_n,\theta_{n-1})\right)
    \label{eq:inequations_L}
\end{equation}
where
\begin{equation}
    f_1(\theta_n,\theta_{n-1})=\frac{k_n\cos\theta_n-k_{n-1}\cos\theta_{n-1}}{k_n+k_{n-1}}
\end{equation}
\begin{equation}
  f_2(\theta_n,\theta_{n-1})=\frac{k_n\cos\theta_n-k_{n-1}\cos\theta_{n-1}}{k_n-k_{n-1}}
\end{equation}

As mentioned above, the maximum radio intensity of the electrostatic Langmuir waves occurs about $\theta_L=0^{\circ}$ and $180^{\circ}$ and of the harmonic emission $H$ takes place about $\theta_2=45^{\circ},135^{\circ},225^{\circ}$ and $315^{\circ}$ respectively, then radiation angle $\theta_3$, at which the third emission $H_3$ reaches its maximum radiation intensity, can be solved based on \myrefeq{eq:inequations_L}. For example, for $\theta_L=0^{\circ}$ and $\theta_2=45^{\circ}$, $\theta_3$ lies within range $0-35^{\circ}$. \myreffig{fig:EM_PE_3D}(d) shows for $H_3$ power spectrum within angle range $0-35^{\circ}$ are more tense than that within angle range $ 35-90^{\circ}$ in the first quadrant in the $k_{\parallel}-k_{\perp}$ plane, for simplicity without loss of generality, we choose $\theta_3=20^{\circ}$. In the same vein, the radiation angles for $H_3$ and $H_4$ solved from \myrefeq{eq:inequations_L} are summarized in \myreftab{tab:radiation_angle}.

\begin{deluxetable*}{cccl}
  \tablewidth{0pt}
  \tablenum{3}
  \tablecaption{Angle location of the strongest emission for transverse harmonics $H_n$. The angle location of the strongest intensity for electrostatic Langmuir waves is taken as $\theta_L=0^{\circ}$ and $180^{\circ}$.\label{tab:radiation_angle}}
  \tablehead{
    $H_n$ & $\theta_L$ & $\theta_{n-1}$  & $\theta_{n}$\\
  }
  \startdata
  $F$             &                &                 & $90^{\circ},270^{\circ}$\\
  \cline{1-4}
  $H$             &                &                 & $45^{\circ},135^{\circ},225^{\circ},315^{\circ}$\\
  \cline{1-4}
  \colhead{$H_3$} &  $0^{\circ}$   & $45^{\circ}$    & $0-35^{\circ}$\\
  \colhead{}      &  $0^{\circ}$   & $315^{\circ}$   & $325-360^{\circ}$\\
  \colhead{}      &  $180^{\circ}$ & $135^{\circ}$   & $145-180^{\circ}$\\
  \colhead{}      &  $180^{\circ}$ & $225^{\circ}$   & $180-215^{\circ}$\\
  \hline
  \colhead{$H_4$} &  $0^{\circ}$   & $20^{\circ}$    & $0-17^{\circ}$\\
  \colhead{}      &  $0^{\circ}$   & $340^{\circ}$   & $343-360^{\circ}$\\
  \colhead{}      &  $180^{\circ}$ & $160^{\circ}$   & $163-180^{\circ}$\\
  \colhead{}      &  $180^{\circ}$ & $200^{\circ}$   & $180-197^{\circ}$\\
  \hline
  \colhead{$H_5$} &  $0^{\circ}$   & $10^{\circ}$    & $0-9^{\circ}$\\
  \colhead{}      &  $0^{\circ}$   & $350^{\circ}$   & $351-360^{\circ}$\\
  \colhead{}      &  $180^{\circ}$ & $170^{\circ}$   & $171-180^{\circ}$\\
  \colhead{}      &  $180^{\circ}$ & $190^{\circ}$   & $180-189^{\circ}$\\
  \enddata
\end{deluxetable*}

%%%%%%%%%%%%%%%%%%%%%%%%%%%%%%%%%%%%%%%%%%%%%%%%%%%%%%%
\bibliographystyle{aasjournal}
%\bibliography{references}

\end{document}